\documentclass[]{interact}

\usepackage{epstopdf}
\usepackage[caption=false]{subfig}

\usepackage[numbers,sort&compress]{natbib}
\bibpunct[, ]{[}{]}{,}{n}{,}{,}
\renewcommand\bibfont{\fontsize{10}{12}\selectfont}

\theoremstyle{plain}

\theoremstyle{definition}

\theoremstyle{remark}

\usepackage{multirow}

\begin{document}


\title{Flamelet Connection to  Turbulence Kinetic Energy Dissipation Rate}

\author{
\name{William A. Sirignano\textsuperscript{a}\thanks{CONTACT William A. Sirignano. Email: william.a.sirignano@uci.edu}, Wes Hellwig\textsuperscript{a} and Sylvain L. Walsh\textsuperscript{a}}
\affil{\textsuperscript{a}University of California, Irvine, Irvine, California, USA}
}

\maketitle

\begin{abstract}
The turbulence kinetic energy dissipation rate $\epsilon$, from a turbulent combustion computation using either Reynolds-averaged Navier-Stokes (RANS) or large-eddy simulation (LES), is proposed for closure with a sub-grid non-premixed flamelet model. The intentions are to avoid the creation of artificial tracking or progress variables and to relate accurately the physics of turbulent non-premixed combustion at the resolved length scales to the small-scale physics where the mixing and chemical reactions occur. The analysis addresses the relations between $\epsilon$ and the strain rate, vorticity, viscous dissipation rate, scalar gradients, scalar dissipation rate, and burning rate at the smallest turbulence length scales where diffusion-controlled burning is faster than at larger length scales and thereby dominant. The imposed strain rate and vorticity on these smallest eddies are determined from the kinetic energy dissipation rate. Thus, an $\epsilon$ value at a specific time and location determines the two mechanical constraints (vorticity and strain rate) on the inflow to the counterflow flamelet. $\epsilon$ affects the sign of the Laplacian of pressure, which must be negative to allow the existence of the counterflow. Using different flamelet models, with and without vorticity, different results for maximum flamelet temperature, integrated flamelet  burning rate, and maximum flamelet scalar dissipation rate are obtained. Flamelet models that consider the centrifugal effect of vorticity produce substantial enhancements in the accuracy and completeness of information for a turbulent combustion computation. $\epsilon$ may be used as a tracking variable that connects the sub-grid flamelet model to resolved-scale RANS or LES computations. ARTICLE WORD COUNT: 8060 words.
\end{abstract}


Large-eddy simulations (LES) or Reynolds-averaged Navier-Stokes (RANS) are used for turbulent combustion computations where the smallest scales  cannot be resolved.  The smaller scales are filtered via integration over a window size commensurate with the computational mesh size, thereby allowing affordable computations. The essential, rate-controlling, physical and chemical processes that occur on scales smaller than the filter size must be described by flamelet closure models.  The flamelet models must handle multi-species, multi-step oxidation kinetics without requiring small time steps during the solution of the resolved-scale fluid dynamics. The focus here is on the types of turbulent flames found in the shear-driven flows of practical combustors. Thereby, we address flamelets as originally described by Williams \cite{Williams1975}, namely ``highly sheared small diffusion flamelets" and ``forming a turbulent flame brush which appears on the average to fill"  the flow domain.

The use of the turbulence kinetic energy dissipation rates to connect large-length-scale behavior to the smallest scale behavior is well established for the mechanics of turbulent flow \cite{Pope2000}. It has been used for many decades to relate resolved-scale behavior in Reynolds-averaged Navier-Stokes (RANS) solvers and large-eddy simulations (LES) to sub-grid behavior. Here, with a clear connection between scalar gradients on the smallest scale and the resolved-scale mechanics, we propose a new method to relate the small scales where diffusion-controlled combustion occurs to the resolved-scale flow. Focus will be placed on the two-way coupling, including the feedback of the heat release rate from the small-length scales to the resolved scale.

In its most basic sense, the flamelet model quantifies the unresolved fluid-chemistry interaction occurring at each RANS or LES node or cell. The fluid-chemistry interaction will be determined by a set of governing equations which are subject to two types of constraints: 1) mechanical constraints, indicative of velocity gradients and thus residence time; and 2) thermodynamic constraints, which are indicative of mixture, temperature, and pressure. These constraints on the flamelet model must be a function of known RANS or LES quantities for the overall CFD framework to close the system. In this way, the coupling between RANS or LES and the flamelet model parallels the classical closure approach used to relate kinetic energy on the averaged or resolved scale to the sub-grid dissipation of that energy. We specifically target the mechanical constraints of the Rotational Flamelet Model (RFM) \cite{Sirignano2022a,Sirignano2022b,Hellwig2025a,Hellwig2025b}; however, the procedure laid out in this work can also be applied to other flamelet models. This work does not address a new coupling for thermodynamic or chemical  constraints.

The formulation here does not input scalar dissipation rate or progress variable into the subgrid flamelet model and does not use a probability density function but rather provides an alternative coupling procedure with a differing foundation. For LES, we especially aim for detached-eddy simulations where, like RANS, resolutions of turbulence kinetic energy $k$ and turbulence kinetic energy dissipation rate $\epsilon$ are determined to describe the flow below the filtered length scale. The coupling for the mechanical constraints established here is conditionally formulated on the ability of the RANS or LES code to quantify $\epsilon$.

The method builds upon a few established concepts in the science of turbulence. The cascade of turbulence kinetic energy through the length scales has mechanical energy being transferred from larger eddies to sequentially smaller eddies with insignificant dissipation until the range of smallest eddy sizes is reached. Although there is little turbulence kinetic energy in these smallest (Kolmogorov and near-Kolmogorov) scales, the viscous dissipation rate dominates there because they contain the largest velocity derivatives. Thereby, these scales also contain the largest values of vorticity (or enstrophy) and normal strain rates in the spectrum. Since viscosity is the diffusion of momentum or vorticity, we expect, for turbulent flows with near-unity values of Prandtl number and Schmidt number, scalar gradients and mixing rates for heat and mass will be dominate on those same length scales, where kinetic energy dissipation occurs.  Furthermore, since the smallest eddies have low values of kinetic energy, compared to larger length scales in the spectrum (as shown by Figure 6.22 of Pope 2000 \cite{Pope2000}), their impact on eddy diffusivity or flow curvature on larger scales is not significant. However, molecular mixing and diffusion-controlled combustion will dominate on this scale. On average, we assume the distance travelled by the gaseous material in the turbulence cascade from the end of the computational resolution to the dissipation completion is small. However, since the distance between two material points can decrease due to sever straining, scalar gradients will increase as length scale decreases.

Some valuable guidance on the relation between scalar mixing and viscous dissipation is provided by the work of Yeung, Donvis, and Sreenivasan (YDS) \cite{yeung_high-reynolds-number_2005}. Following them, $\delta u$ is the velocity increment over a distance $\delta r$ in the same direction as the velocity vector. $\delta \phi$  is the conserved scalar increment in that same direction and distance. $\phi$ is non-dimensional and $O(1)$ over the flow. From Equations (1, 3) and Figures (1, 3) of the YDS paper, $\left(\delta u\right)^3/\left(\epsilon \delta r\right)=4/5$, where $\epsilon$ is the viscosity-based turbulence kinetic energy dissipation rate. The scalar dissipation rate is $\chi = \left(3/2\right)\left(\delta \phi\right)^2 \delta u/\delta r$. Based on this definition, we also expect $\chi = 2 D \left(\delta \phi/ \delta r\right)^2$ where $D$ is the thermal or mass diffusivity. Now, setting  $\delta u \delta r / \nu = 1$ (for Kolmogorov scales where $\nu$ is the kinematic viscosity), we have  $\epsilon/\nu = \left(5/4\right) \left(\delta u / \delta r\right)^2$ and $\delta u / \delta r = \left(2/3\right) \chi /\left(\delta \phi\right)^2$. Thus, $\chi = 3(\delta \phi)^2\left(\epsilon/(5\nu)\right)^{\frac{1}{2}}$ where $\delta \phi$ is a nondimensional variable smaller than $O(1)$. Also, $\chi \sim \delta u / \delta r$. That is, strain rate determines scalar dissipation rate on this scale.

A brief literature review of the Classical Flamelet Model (CFM) and its mechanical coupling procedures is presented in Section \ref{CFM}. An overview of the Rotational Flamelet Model (RFM) is presented in Section \ref{RFM}. An overview of key relations for the RFM is given in Section \ref{basics}. The description of the velocity divergence and curl in the RFM rotating reference frame is discussed in Section \ref{dissenst} while the viscous dissipation, enstrophy, and kinetic energy in that frame are discussed in Sections \ref{dissenst} and \ref{ke}. Scaling of velocity gradients between the resolved scale and flamelet scale are analyzed in Section \ref{grad}; specific use of and connection of the flamelet inflow with turbulence kinetic energy dissipation rate is addressed alongside results of calculations for several flamelet models in Section \ref{connect} and Conclusions are given in Section \ref{conclusions}. While the science and key relations are provided for two-way flamelet-flow connection, the computational application with  a computation for a turbulent combustion field is left for future study.

\section{Classical Flamelet Model and Coupling Procedure for Mechanical Constraints} \label{CFM}

It is critical to understand the laminar mixing and combustion that commonly occur within the smallest turbulent eddies. These laminar flamelet sub-domains experience significant strain of all types: shear, tensile, and compressive. Notable works exist here for either counterflows with only normal strain or simple vortex structures without stretching in two-dimensions or axisymmetry and sometimes with a constant-density approximation implied through an assumption of uniform nomral strain rate. See \cite{Peters,Pierce,Linan,Marble,Karagozian,Cetegen1,Cetegen2}.  An interesting review of the early flamelet theory is given by \cite{Williams2000} and a recent review is given by \cite{DOMINGO2025}. Flamelet studies have generally focused on either premixed or non-premixed flames, although some recent works have adopted a unifying approach to premixed, non-premixed, and multi-branched flames, referred to as multi-modal combustion.

Vorticity interaction with the flame has not been directly considered in most models \citep{Peters, Pierce, Linan,Williams2000}. Two-dimensional planar and axisymmetric counterflow configurations are generally the foundation for a flamelet model. Local conversion to a coordinate system based on the principal strain-rate directions typically provides the counterflow configuration. Furthermore, the quasi-steady counterflow can be analyzed by ordinary differential equations because the dependence on the transverse coordinate is either constant or linear, depending on the variable. The mixture fraction has been used widely as an independent variable to display non-premixed flamelet scalar variations; this cannot be useful for premixed flames. It has been shown by \cite{Sirignano2021a} that any conserved scalar can serve well as an independent variable to present scalar results for nonpremixed and multi-branched flamelets.

The classical counterflow treatment \citep{Linan, Peters, Bilger}, hereafter refereed to as the ``Classical Flamelet Model'' (CFM), solves the flamelet equations in mixture fraction space, yielding laminar reactive scalars $\psi_i=\psi_i(Z,\chi)$. Once these are obtained, a coupling procedure is executed whereby resolved scale RANS or LES quantities indicate which of the many $\chi$ (mechanical constraint) profiles to use for the given resolved-scale node or cell. A similar coupling must be given for the thermodynamic state, but we do not discuss this in detail here. The mechanical coupling is typically performed in one of two ways using the presumed PDF approach: 1) use $\chi$ directly in a PDF (This method involves only the two stable branches of the S-curve and does not involve the unstable branch as possible flamelet states); or 2) bijectivley map $\chi$ to a monotonic Flamelet Progress Variable (FPV) $C$ \cite{Pierce} and use that variable along with a presumed PDF approach (this method includes all branches of the S-curve as realizable flamelet states).


In a presumed PDF approach, the Favre mean $\widetilde{\chi}$ is an input to the PDF and must be modeled on the resolved scale. This is commonly done as $\widetilde{\chi} = C_{\chi}(\widetilde{\epsilon}/\widetilde{k})\widetilde{Z''^2}$ where $C_{\chi}$ is the proportionality constant between turbulence and scalar time-scales. $C_{\chi}$ is typically set to 2.0; however, this value was determined specifically with an inert jet of methane \cite{janicka_prediction_1982}. A modeled equation can also be used, see \cite{Elghobashi1977,ElghobashiLaunder1983,NewmanLaunderLumley1981,BorghiDutoya1979}. 

The purpose of the PDF for $\chi$ is to couple the LES or RANS quantities to the flamelet quantities. Flamelets exist on length scales smaller than the LES filter scale and below the mean produced by RANS; thus, the flamelet should not be directly subjected to the LES or RANS quantities, but rather to the instantaneous flow field that is filtered or averaged away by LES and RANS. Of course, the actual instantaneous flow field is unknown. So, a model of the turbulent fluctuations is used. The log-normal PDF $P(\chi)$ is one such model which, given a mean $\widetilde{\chi}$, describes the probability of instantaneous scalar dissipation rate occurring in the range $0 < \chi < \infty$. Since the flamelet equations are solved for the same range of $\chi$, convolution between $\psi_i(Z,\chi)$ and $P(\chi)$ is claimed to characterize the fluctuations and intermittency of instantaneous small-scale eddies and their effect on combustion. The log-normal nature of $P(\chi)$ stems from Kolmogorov's third hypothesis and has been shown to be in good agreement with experiments and numerical data \cite{SreenivasanAntoniaDanh1977,SreenivasanAntonia1977,Kerstein1984,Dahm1989} provided the length scale observed was between $\kappa << r << L$ where $\kappa$ is the Kolmogorov length scale and L is the integral length scale. Sreenivasan et al. \cite{SreenivasanAntoniaDanh1977} observe lognormality of the PDF between $5\kappa < r < 150\kappa$. Other authors have used a delta function for $P(\chi)$ centered on $\widetilde{\chi}$ which completely ignores fluctuations. Sanders and Lamers \cite{sanders_modeling_1994} match experiments better in the prediction of turbulent diffusion flame liftoff height, using  a Gaussian PDF for strain rate of the smallest length scales, rather than a log normal PDF for scalar dissipation rate. The mean and variance of that strain rate PDF are proportional to the square root of the turbulence kinetic energy dissipation rate in their RANS formulation. That dependence relates to our analysis and results  in the following sections.

Regardless of the PDF used, the present authors are concerned about a mechanical coupling using $\chi$ for two related reasons. First, it requires a presumed profile $\chi(Z)$ to solve the CFM equations which invokes multiple simplifying assumptions and thus far has not included the effect of Kolmogorov vorticity at the smallest scales \cite{Sirignano2022a,Sirignano2022b,Hellwig2025a,Hellwig2025b}. Furthermore, the definition of scalar dissipation rate of a conserved scalar requires equality among the various species diffusivities. We do not wish to invoke these various assumptions. Second, we hypothesize that the diffusion layer held by a Kolmogorov eddy and other small scale eddies see the influence of mechanical constraints as far-field conditions. The flame-local structure is then determined through governing equations subjected to three far-field strain rates and vorticity. Hellwig et al. \cite{Hellwig2025a} shows that inclusion of these four parameters creates a multi-valued correspondence between $S^*$ and $\chi_{st}$ whereas the assumed profile $\chi(Z)$ produces a one-to-one correspondence. Therefore, it is preferred to use a formulation considering the asymptotic strain rate, $S^*$, as well as the other two transverse strain rates and vorticity, rather than an assumed profile for $\chi$ based on $S^*$ alone. We seek a coupling procedure that can quantify these mechanical constraints. Furthermore, the use of a PDF for $\chi$ introduces an inconsistency with the scalar dissipation rate term in the resolved-scale equation for scalar variation (the variance of the mixture fraction in the CFM approach), namely, $\widetilde{\chi} = C_{\chi}(\widetilde{\epsilon}/\widetilde{k})\widetilde{Z''^2}$ where the characteristic time for that rate is assumed to be proportional to the characteristic time for energy dissipation rate. There are implicitly two scalar dissipation rate definitions involved.

These concerns are exacerbated when adopting a progress variable (FPV) framework. In addition to requiring the same assumptions about scalar dissipation rate profiles and the modeling of the scalar dissipation rate on the resolved scale, the FPV approach introduces a more fundamental issue: a loss of the mechanical constraint fidelity. Walsh et al. \cite{walsh_turbulent_2025} found that once flamelet solutions are mapped from scalar dissipation rate (i.e., strain rate) to progress variable the, intrinsic relationship between strain rate and flamelet structure is no longer preserved in the resolved-scale solution. This decoupling arises because the progress variable evolves according to its own transport equation at the resolved scale, independent of the local strain rate, despite the fact that the strain rate governs the underlying flamelet structure in the pretabulated solutions. As a consequence, the resolved-scale model cannot retain the correlation between strain rate and the progress variable values that exists at the subgrid flamelet level. It has been shown \cite{walsh_performance_2025} that this decoupling leads to the preferred selection of equilibrium flamelet (low strain solutions) solutions in regions of high strain, resulting in nonphysical heat release rates. So here, the mechanical constraint becomes non-strain rate informed.

The theory presented in this paper is based on the hypothesis that the smallest eddies, those on or around the Kolmogorov scale and which have the highest strain rates and shortest time scales, have leading order effects on combustion since they produce the highest heat release rates. Furthermore, these eddies on the near-Kolmogorov scales are the only eddies which are laminar and thus the only sensible location for a laminar flamelet. Therefore, the goal of this work is to derive a coupling procedure which approximates the strain rate and enstrophy $\omega^2$ magnitude of these particular eddies. We do not attempt to predict stochastically the variation in strain rate due to intermittency. Instead, we take a spatial average over a Kolmogorov volume $\kappa^3$ to approximate the average mechanical constraint experienced by that Kolmogorov eddy. A temporal average needed to account for intermittency will be applied through a coefficient. In other words, the fluctuations of Kolmogorov eddies will be averaged out.

We present here an alternative coupling procedure which connects sub-grid strain rate and thereby sub-grid scalar behavior directly to the turbulence kinetic energy dissipation rate $\epsilon$. Specifically, we connect the sub-grid scalar dissipation rate to the resolved scale whereby $\epsilon$ is used to quantify the strain rate $S^*$ and the vorticity magnitude $\omega$, that a laminar counterflow flamelet existing on the Kolmogorov scale, would experience without the use of a PDF approach. These asymptotic quantities are grounded firmly in turbulence cascade theory and may thus offer a more physically-sound coupling of the mechanical constraints between scales for the specific viewpoint of the flamelet model as a Kolmogorov-scale model. Although the approach is developed with the RFM in mind, as shown in Table \ref{tab}, it may also be applied to the CFM approach, as will become evident in Section \ref{connect}.


\section{Rotational Flamelet Model} \label{RFM}

The Rotational Flamelet Model (RFM) is a subgrid, Kolmogorov-scale flamelet model originally created by Sirignano \cite{Sirignano2022a}. It uses a 3D, non-Newtonian reference frame rotating at half of the vorticity magnitude such that shear strain is removed from the inflows; i.e., three principal strain rates exist. The major tensile strain rate is $S_{1}^*=S_1S^*$, the major compressive strain rate is $S^*$, and the intermediate strain rate, which may be tensile or compressive, is $S_{2}^* = S_2S^*$. Scalar gradients are aligned with the major compressive strain rate direction. Vorticity and the stretching of vorticity align with the major extensional direction. Through the coordinate transformation, a similar solution is found for the governing equations, reducing it to a system of ordinary differential equations (ODEs). The vorticity is uniform across the small flamelet domain. This is justified because vorticity maximizes towards the center of the eddy creating a zero gradient of vorticity locally. Sirignano considered both a situation where the intermediate strain rate was tensile \cite{Sirignano2022a} where a stretched vortex sheet or layer is created, and a stretched vortex tube situation with inward swirl created by two compressive-strain directions and the extensional direction aligned with the vorticity vector \cite{Sirignano2022b}. This tube presented an analog to the classical Burgers stretched vortex \citep{Burgers1948, Rott}. In both models, vorticity is shown to create a centrifugal force on the sub-grid counterflow that modifies the molecular transport rates and burning rate. This has consequence on the flow rate through the flame zone and the burning rate. It thereby modifies flammability limits. These initial demonstrations of the RFM were done with Fickian diffusion, unity Lewis numbers, and one-step chemical kinetics.

Hellwig et al. \cite{Hellwig2025b} improved the rotational flamelet model by including detailed kinetics, multicomponent transport, and variable thermophysical properties with a more complete examination of flammability limits. The quasi-steady assumption commonly used in previous flamelet models was retained here for the flamelet. Hellwig et al. \cite{Hellwig2025a} later substituted a Fickian diffusion formulation, setting the diffusivities of all species equal to the thermal diffusivity in order to define scalar dissipation rate in the RFM and showed that vorticity and varying transverse strain rates create a multi-valued correspondence between the asymptotic strain rate $S^*$ and the flame-local scalar dissipation rate. This work was conducted because the scalar dissipation rate is given importance in the CFM and FPV methods, which neglected vorticity 3D effects. The RFM has the direct input of both strain rate and vorticity. For both models,  a given maximum temperature or burning rate correlates with a specific scalar dissipation rate (SDR) in the reaction zone. However, different imposed values of compressive strain rate and vorticity impose that same value of SDR. Both \cite{Hellwig2025a,Hellwig2025b} showed fundamental results consistent with the original works of Sirignano \cite{Sirignano2022a,Sirignano2022b}, in that the combination of varying transverse strain rates and vorticity modify the local flame structure and burning rate. 

Scaling laws have been presented by \citep{Sirignano2022a} for relating strain rates and vorticity at the sub-grid level to those quantities at the resolved-flow level for coupling with large-eddy simulations or the time-averaged mean-flow level for Reynolds-averaged flows. The time-averaged behavior of a simple turbulent flow is resolved with coupling to the RFM. Specifically, a two-dimensional, multicomponent, time-averaged  planar shear layer with variable density and energy release is employed using a mixing-length description for the eddy viscosity.

Useful background information relevant to turbulent non-premixed combustion is given by References \cite{Elghobashi1977,Nomura1992,Elghobashi1974, Nomura1993, Boratav1996, Boratav1998, Ashurstetal, Dahm}. Background for premixed turbulent combustion is found with References \cite{Driscoll, Steinbergetal, Steinberg, Hamlington, Swaminathan}.
  
Here, we address the use of turbulence kinetic energy dissipation rate $\epsilon$ to connect the RANS resolved-scale or LES filtered-scale for turbulent, globally non-premixed combustion to the sub-grid flamelet model.  The quantity $\epsilon$ is a well-established parameter in turbulence theory \citep{JonesLaunder, Pope2000} and will be used as a coupling variable to determine the parameters (vorticity and strain rate) at the smallest scale of turbulence that are constraints for the flamelet model.  Specifically, we address here the inputs to the RFM \cite{Sirignano2022a,Hellwig2025a,Hellwig2025b,Sirignano2022b}. We also make comparisons with CFM results generated by Flamemaster \citep{flamemaster}, which are based on references \cite{Peters, Pitsch1, Pitsch2}.

\section{ Basic Relations for Rotational Flamelet} \label{basics}

\begin{figure}[h]
  \centering
  \includegraphics[width=3.25in]{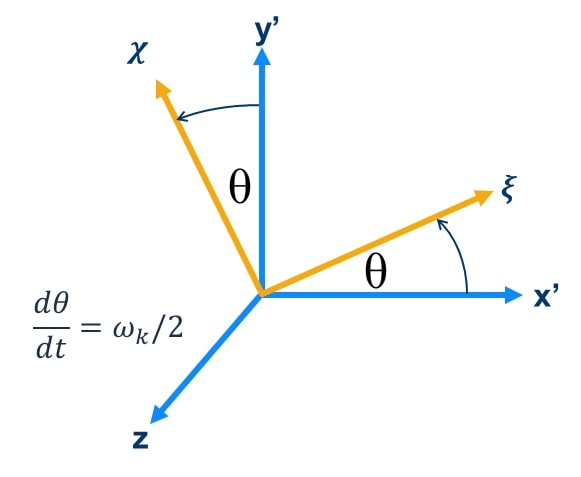}
  \caption{Transformation to $\xi, \chi, z'$ rotating coordinate system from $x', y', z'$ Newtonian system. $\theta$ increases in the counterclockwise direction.   }
  \label{Coordinate}
\end{figure}

Consider the rotational flamelet transformation from the non-rotating coordinate system with space coordinates $x, y, z$ and corresponding velocity components $u,v, w$ to the system rotating about the $z$-axis through the origin with the coordinates $\xi, \chi, z$ and velocity components $u_{\xi}, u_{\chi}, w$ \citep{Sirignano2022a}. Note that in this section, the symbol $\chi$ refers to the rotating-frame coordinate, not to the scalar dissipation rate as discussed in previous sections. The transformation displayed in Figure \ref{Coordinate} is made from the Newtonian frame with rotating material (due to vorticity) to a rotating, non-Newtonian frame.  The vorticity direction is the $z$ direction in an orthogonal framework. Any $x, y$ plane contains the directions of scalar gradients, major principal axis for compressive strain, and a principal axis for tensile strain.  Note that the $x, y, z$ directions are not correlated with coordinates on the resolved scale. $\omega$ is the vorticity magnitude on this sub-grid  scale. $x, y, z$ are transformed to $\xi, \chi, z$ wherein the material rotation is removed from the $\xi, \chi$ plane by having it rotate at angular velocity $d\theta/dt = \omega/2$ relative to $x, y$. Here, $\theta$ is the angle between the $x $ and $\xi$ axes and simultaneously the angle between the $y$ and $\chi$ axes. The sub-grid domain is sufficiently small to consider a uniform value of $\omega$ across it.

The dependent and independent variables will be presented in dimensional form unless noted otherwise. The relations between the velocity components and spatial coordinates are given as
\begin{eqnarray}
\xi &=& x cos \theta +y \sin \theta  \; ;  \; \chi = y cos \theta - x sin \theta    \\
u_{\xi} &=& u cos \theta + v sin \theta + \chi\frac{\omega}{2}  \;  ;  \;  u_{\chi} = v cos \theta - u sin \theta - \xi\frac{\omega}{2}     \\
\frac{\partial u}{\partial x}  & = &   \frac{\partial u}{\partial \xi}  cos \theta -   \frac{\partial u}{\partial \chi} sin \theta \;   ;  \;   \frac{\partial u}{\partial y}   =    \frac{\partial u}{\partial \xi}  sin \theta +   \frac{\partial u}{\partial \chi} cos \theta    \\
\frac{\partial v}{\partial x}  & = &   \frac{\partial v}{\partial \xi}  cos \theta -   \frac{\partial v}{\partial \chi} sin \theta \;   ;  \;   \frac{\partial v}{\partial y}   =    \frac{\partial v}{\partial \xi}  sin \theta +   \frac{\partial v}{\partial \chi} cos \theta   
\label{1}
\end{eqnarray}

From Equation (1), it follows that $\sqrt{x^2 + y^2} = \sqrt{\xi^2 + \chi^2}$. Thus, the units of length are equal in both reference frames. 

Differentiation of Equations (2), together with constraints that $u_{\xi}$ is independent of $\chi$ and $u_{\chi}$ is independent of $\xi$, yields
\begin{eqnarray}
\frac{\partial u_{\xi}}{\partial \xi}  & = &   \frac{\partial u}{\partial \xi}  cos \theta +   \frac{\partial v}{\partial \xi} sin \theta \;  \; ; \; \;   
\frac{\partial u_{\xi}}{\partial \chi} = 0=   \frac{\partial u} {\partial \chi}  cos \theta +   \frac{\partial v}{\partial \chi} sin \theta + \frac{\omega}{2}  \\ 
 \frac{\partial u_{\chi}}{\partial \chi}  & = &   \frac{\partial v}{\partial \chi}  cos \theta -  \frac{\partial u}{\partial \chi} sin \theta  \; ; \;
\frac{\partial u_{\chi}}{\partial \xi}   =  0=  
\frac{\partial v}{\partial \xi}  cos \theta
-   \frac{\partial u}{\partial \xi} sin \theta   -\frac{\omega}{2}
\end{eqnarray}
 Equations (5) and (6) now can yield
\begin{eqnarray}
\frac{\partial u_{\xi}}{\partial \xi} &=& 
\frac{\partial u}{\partial x} +
\Big(\frac{\partial u}{\partial y} + \frac{\omega}{2}\Big) tan \theta  \\
\frac{\partial u_{\chi}}{\partial \chi} &=& 
\frac{\partial v}{\partial y} -
\Big(\frac{\partial v}{\partial x} - \frac{\omega}{2}\Big) tan \theta 
 \end{eqnarray}

Now, we focus on the inflow on either side of the counterflow induced by a compressive normal strain rate imposed on the rotating mass of fluid. Asymptotically with distance from the center of rotation, the inflow has uniform density, although exothermic reaction in the center of the rotating fluid will cause variable density locally.  The divergence of velocity in either coordinate system will be zero for the incoming flow. 
The inflow should be independent of $\theta$ position. That occurs if $\partial v / \partial x = - \partial u/ \partial y = \omega/2$. This implies that the shear strain rate $S_{xy} = [\partial u/ \partial y + \partial v / \partial x]/2 =0$. Since $w$ depends only on $z$ and $u$ and $v$ are independent of $z$ for the inflow, $S_{xz} =0$, $S_{yz}=0$ as well. Thereby, only normal strain rate occurs in the inflow for either coordinate system.

With the aim towards establishing scaling relations, the flamelet-related material  is applied to incoming counterflow at some distance from the reaction zone. There, density, other scalar properties, and strain rates are considered as uniform.  Velocity gradients and scalar gradients will differ from values at the filtered scales of the resolved flow. Our scaling from resolved-scale values for RANS or LES will impose boundary conditions on the flamelet but will not prescribe directly the values in or near the reaction zone; in this critical manner, the approach differs from the Flamelet Progress Variable method \citep{Pierce}.

Given the relations for the vorticity in the original reference frame and the transformed frame,  we have 
\begin{eqnarray}
\frac{\partial v}{\partial x} - \frac{\partial u}{\partial y }  = \omega
\;\; ; \;\; 
\frac{\partial u_{\chi}}{\partial \xi} - \frac{\partial u_{\xi}}{\partial \chi }  = 0
\end{eqnarray}    
The dilation remains identical in both frames; thus,
it follows that the divergence of velocity is unchanged in value. Namely,
\begin{eqnarray}
\frac{\partial u_{\chi}}{\partial \chi} + \frac{\partial u_{\xi}}{\partial \xi} + \frac{\partial w}{\partial z} = 
\frac{\partial u}{\partial x} + \frac{\partial v}{\partial y}  + \frac{\partial w}{\partial z}
\label{11}
\end{eqnarray}
Furthermore, the divergence will be zero for the incoming flow; although density variation will occur in the flamelet interior due to exothermic reaction.

For flow into the flamelet counterflow at a distance from the interface of the two incoming streams,  $u_{\xi}$ is only a function of $\xi$ and $u_{\chi}$ is only dependent on
$\chi$, and the  velocity derivatives $\partial u_{\xi}/\partial \chi$ and  $\partial u_{\chi}/\partial \xi$          are each zero. 
That is, the vorticity does not appear explicitly for the inflow in the description within the rotating frame. It can have the important implicit effect through the centrifugal acceleration. In the flamelet interior core with varying density, velocity derivatives indicating shear strain and vorticity can develop because $u_{\xi}(\xi, \chi)$ and $w(\chi, z)$ can appear. However, since $u_{\xi}$ is antisymmetric in $\xi$ and $w$ is antisymmetric in $z$, the circulations for the regions with total vorticity in the $\xi$- or $\chi$-directions or added vorticity in the $z$-direction  will each be zero.

\section{Viscous Dissipation and Enstrophy} \label{dissenst}

Consider first a general viscous flow configuration. Later, we will specifically address our flamelet-related configuration. The vorticity vector is the curl of the velocity, which written in tensor is $\omega_k = \epsilon_{ijk} \partial u_j/\partial x_i$ where    $\epsilon _{ijk}$  is the Levi-Civita symbol. The enstrophy is defined as the dot product of the vorticity vector with itself: $\vec{\omega}\cdot \vec{\omega} = \omega^2$.  Thus, in the original coordinates, 
\begin{eqnarray}
\vec{\omega} &=& (\frac{\partial w}{\partial y} - \frac{\partial v}{\partial z},    \frac{\partial u}{\partial z} - \frac{\partial w}{\partial x},     \frac{\partial v}{\partial x} - \frac{\partial u}{\partial y})  \;\; ; \;\; 
\vec{\omega}\cdot \vec{\omega} = \omega^2\\
\omega^2   &=& \Big(\frac{\partial u}{\partial y}\Big)^2 + \Big(\frac{\partial u}{\partial z}\Big)^2 + \Big(\frac{\partial v}{\partial x}\Big)^2 +\Big(\frac{\partial v}{\partial z}\Big)^2 + 
\Big(\frac{\partial w}{\partial x}\Big)^2 +\Big(\frac{\partial w}{\partial y}\Big)^2  \nonumber   \\
&&-  2  \Big(\frac{\partial v}{\partial x}\frac{\partial u}{\partial y} + \frac{\partial u}{\partial z}\frac{\partial w}{\partial x}   + \frac{\partial v}{\partial z}\frac{\partial w}{\partial y} \Big)          
\end{eqnarray}

For a Newtonian fluid with dynamic coefficient of viscosity $\mu$, application of Stokes hypothesis, and use of tensor notation, the viscous stress $\tau_{ij}$ and the viscous dissipation rate $\Phi$ are given as follows. 
\begin{eqnarray}
\tau_{ij} &=& \mu\Big(\frac{\partial u_i}{\partial x_j} +  \frac{\partial u_j}{\partial x_i} 
-\frac{2}{3} \delta_{ij}\frac{\partial u_k}{\partial x_k} \Big) \label{15}  \\
\Phi& =& \tau_{ij}\frac{\partial u_i}{\partial x_j} = \mu\Big(\frac{\partial u_i}{\partial x_j}\frac{\partial u_i}{\partial x_j} +  \frac{\partial u_j}{\partial x_i}\frac{\partial u_i}{\partial x_j} 
-\frac{2}{3} \Big[\frac{\partial u_k}{\partial x_k} \Big]^2\Big)  \label{16} \\
\frac{\Phi}{\mu}  &=& \frac{4}{3}\Big[  \Big(\frac{\partial u}{\partial x} \Big)^2  + \Big(\frac{\partial v}{\partial y} \Big)^2 + \Big(\frac{\partial w}{\partial z} \Big)^2  \Big]  +  \Big(\frac{\partial u}{\partial y} \Big)^2  + \Big(\frac{\partial v}{\partial x} \Big)^2  + 2\frac{\partial u}{\partial y}\frac{\partial v}{\partial x}  \nonumber \\
&& - \frac{4}{3} \Big[ \frac{\partial u}{\partial x}\frac{\partial v}{\partial y}   + \frac{\partial u}{\partial x}\frac{\partial w}{\partial z} +\frac{\partial v}{\partial y}\frac{\partial w}{\partial z}\Big] \nonumber \\
&&+ \Big[\Big(\frac{\partial u}{\partial z} \Big)^2  + \Big(\frac{\partial v}{\partial z} \Big)^2 + \Big(\frac{\partial w}{\partial x} \Big)^2  + \Big(\frac{\partial w}{\partial y} \Big)^2  + 2\frac{\partial u}{\partial z}\frac{\partial w}{\partial x}   + 2\frac{\partial v}{\partial z}\frac{\partial w}{\partial y}  \Big] \label{17}
\end{eqnarray}
In our special flamelet case where $u$ and $v$ are independent of $z$, four terms on the last line of Equation (\ref{17}) become zero-valued. The other two terms on that line are zero for the incoming flow where $w$ depends only on $z$.

With $\omega/2 = \partial v/ \partial x = - \partial u/ \partial y$, as shown above,  Equations (7) and (8) yield
$\partial u/ \partial x =  \partial u_{\xi}/ \partial \xi$  and $\partial v/ \partial y = \partial u_{\chi}/\partial \chi$. Also, the last three terms on the first line of Equation (\ref{17}) become $\omega^2/4 +\omega^2/4 -2\omega^2/4 =0$.  Now, for the inflow, the viscous dissipation can be described by
\begin{eqnarray}
\frac{\Phi}{\mu}  = \frac{4}{3}\Big[  \Big(\frac{\partial u}{\partial x} \Big)^2  + \Big(\frac{\partial v}{\partial y} \Big)^2 + \Big(\frac{\partial w}{\partial z} \Big)^2     
- \frac{\partial u}{\partial x}\frac{\partial v}{\partial y}   - \frac{\partial u}{\partial x}\frac{\partial w}{\partial z} -\frac{\partial v}{\partial y}\frac{\partial w}{\partial z}\Big]   \nonumber \\
 = \frac{4}{3}\Big[\Big(\frac{\partial u_{\xi}}{\partial \xi} \Big)^2  + \Big(\frac{\partial u_{\chi}}{\partial \chi} \Big)^2 + \Big(\frac{\partial w}{\partial z} \Big)^2  - \frac{\partial u_{\xi}}{\partial \xi}\frac{\partial u_{\chi}}{\partial \chi} -\frac{\partial u_{\xi}}{\partial \xi}\frac{\partial w}{\partial z} -\frac{\partial u_{\chi}}{\partial \chi}\frac{\partial w}{\partial z}  \Big]  \label{18}
\end{eqnarray} 

The last relation for $\Phi/\mu$ is identical to the form to be found from direct application of Equation (\ref{17}) using the velocity components and coordinates for the rotating frame.  Essentially, the wheel-like dissipation causes no strain or additional dissipation, yielding the same dissipation rates to be calculated for either frame.

We consider the incoming streams to the flamelet which are divergence free and Equations (\ref{15},\ref{16}, \ref{18}) may be written as
\begin{eqnarray}
\tau_{ij} = \mu\Big(\frac{\partial u_i}{\partial x_j} +  \frac{\partial u_j}{\partial x_i} \Big)
\;\;  ; \;\;
\Phi = \tau_{ij}\frac{\partial u_i}{\partial x_j} = \mu\Big(\frac{\partial u_i}{\partial x_j}\frac{\partial u_i}{\partial x_j} +  \frac{\partial u_j}{\partial x_i}\frac{\partial u_i}{\partial x_j} \Big)   \;\;\; ; \nonumber \\
\frac{\Phi}{\mu}  = 2\Big[ \Big(\frac{\partial u}{\partial x} \Big)^2  + \Big(\frac{\partial v}{\partial y} \Big)^2 + \Big(\frac{\partial w}{\partial z} \Big)^2     
\Big]    = 2\Big[\Big(\frac{\partial u_{\xi}}{\partial \xi} \Big)^2  + \Big(\frac{\partial u_{\chi}}{\partial \chi} \Big)^2 + \Big(\frac{\partial w}{\partial z} \Big)^2 \Big] \label{19}
\end{eqnarray}

If the divergence of the momentum equation is taken for the case of constant density and constant coefficient of viscosity, we obtain with use of relations for $\Phi$ and $\omega^2$
\begin{eqnarray}
\frac{1}{\rho}\frac{\partial^2 p}{\partial x_i^2} &=&  - \frac{\partial}{\partial x_i} \Big(u_j\frac{\partial u_i}{\partial x_j}\Big) + \frac{\mu}{\rho}\frac{\partial^2}{\partial x_j^2}\Big(\frac{\partial u_i}{\partial x_i}\Big) \nonumber \\ &=&
 - \frac{\partial u_j}{\partial x_i} \frac{\partial u_i}{\partial x_j} - u_j\frac{\partial }{\partial x_j}\frac{\partial u_i}{\partial x_i}
 +\frac{\mu}{\rho}\frac{\partial^2}{\partial x_j^2}\Big(\frac{\partial u_i}{\partial x_i}\Big)  \nonumber \\
 &=&   - \frac{\partial u_j}{\partial x_i} \frac{\partial u_i}{\partial x_j}   -0 +0 = 
 - \frac{\partial u_i}{\partial x_j} \frac{\partial u_i}{\partial x_j} + {\omega^2} = \frac{\omega^2}{2} -\frac{\Phi}{2\mu}  \label{20}
\end{eqnarray}
Realize that, in the rotating frame, the centrifugal acceleration is given as $\vec{a} = (\xi \omega^2/4, \chi\omega^2/4)$. Thereby, $\nabla \cdot \vec{a} = \omega^2/2$. Since viscous dissipation maintains the same value in both systems, the implication is that the Laplacian of the pressure is identical in both systems. 
In order to maintain a pressure maximum and thereby a stagnation point, the centrifugal effect must be overcome by the effect of the strain rate; i.e., $\omega^2< A_{ij}A_{ij} $ where $A_{ij} \equiv \partial u_i/\partial x_j$.    Note that $\Phi/\mu$ is independent of the coefficient of viscosity. The balance here relates to strain rate and not viscous action. Actually, the viscous effect was removed by the divergence operation. Still, we use use common terminology that refers to $A_{ij}A_{ij} $ as ``dissipation rate".

Here, we have a strong argument for the importance of the balance of vorticity magnitude and strain-rate magnitude or, more precisely, the balance between dissipation rate and enstrophy. There are many studies on these quantities, especially for incompressible flows \citep{PJohnson,JohnsonWilczek, Yeungetal}. If enstrophy $\omega^2$ becomes too large compared to the dissipation rate divided by dynamic viscosity $\Phi/\mu$, the pressure Laplacian becomes positive, thereby disallowing the flamelet counterflow possibility since the required stagnation point located at a pressure maxima cannot occur.  This vital issue of balance cannot be addressed in a formulation that neglects vorticity and only addresses strain rate. For incompressible turbulent flow, this balance has been addressed by \cite{Yeungetal} who find that the temporal statistics for both quantities present similar magnitudes. They find large values of each parameter tend to occur together with $\Phi/\mu> \omega^2$ for 60-65 percent of the data.   Although the study does not include dilatation, it shows support for the concept of mixing and reaction occurring in rotating counterflows provided by the turbulent eddies. In cases where combustion has mixing as rate-controlling, the higher-strain-rate (and thereby smaller) eddies would provide the highest burning rates (provided the mixing rate does not exceed the reaction rate). 

$\Phi$ gives the kinetic energy dissipation rate per unit volume. Division by density gives $\epsilon$, the kinetic energy dissipation rate per unit mass, commonly used in turbulence modeling. We may write enstrophy as
\begin{eqnarray}
 \omega^2 = 
 \Big[\frac{\partial u_i}{\partial x_j}\frac{\partial u_i}{\partial x_j} 
 -  \frac{\partial u_i}{\partial x_j}\frac{\partial u_j}{\partial x_i} \Big]
 \label{21}
\end{eqnarray}
Then, Equations (\ref{16}) and (\ref{21}) yield

\begin{eqnarray}
\epsilon =   \frac{\Phi}{\rho} =  \nu\Big(2\frac{\partial u_i}{\partial x_j}\frac{\partial u_i}{\partial x_j}
-\frac{2}{3} \Big[\frac{\partial u_k}{\partial x_k} \Big]^2  - \omega^2  \Big)=\nu\left( 2\frac{\partial u_i}{\partial x_j}\frac{\partial u_i}{\partial x_j}-\omega^2\right)
\label{22}
\end{eqnarray}
where $\nu = \mu / \rho$ is the kinematic viscosity.  For the constant-density inflow, we may write 
\begin{eqnarray}
\epsilon = 2 \nu \Big[\Big(\frac{\partial u_{\xi}}{\partial \xi} \Big)^2  + \Big(\frac{\partial u_{\chi}}{\partial \chi} \Big)^2 + \Big(\frac{\partial w}{\partial z} \Big)^2  \Big] 
= 2 \nu[(S^*S_1)^2 + S^{*2} + (S^*S_2)^2] \nonumber \\
=2 \nu S^{*2} [ S_1^2 + 1 + S_2^2] = 4 \nu S^{*2} [ S_1^2 + 1 - S_1]   
\label{23}
\end{eqnarray}
where the definition $S_1 + S_2 =1$ has been entered.   The implication of Equation (\ref{23}) is that if $\epsilon$ were known from the resolved scale, a relation between $S^*$ and $S_1$ is established for the rotational flamelet.  This relation could be useful for the Flamelet Progress Variable model as well. For example, with $S_1 = 1/2$, we find $S^* = \sqrt{\epsilon/(3\nu)}$, its peak value. As $S_1$ varies in the range $0\leq S_1\leq 1$, $S^*$ varies narrowly between $\sqrt{\epsilon/(4\nu)}$ and $\sqrt{\epsilon/(3\nu)}$.  In the range $-1\leq S_1\leq 0$ (with inward swirl), $S^*$ varies more strongly between $\sqrt{\epsilon/(12\nu)}$ and $\sqrt{\epsilon/(3\nu)}$.

Viscous dissipation actually occurs over a range of the smallest scales and not totally at the Kolmogorov scale. Therefore, we have reason to multiply $\epsilon$  in Equations (\ref{22}) and (\ref{23}) by the coefficient $C_{vd}$ where $ C_{vd} <1$. This yields

\begin{equation}
C_{vd}\epsilon = 4\nu S^{*2}[S_1^2+1-S_1]
\label{cvdeps}
\end{equation}

\section{Kinetic Energy} \label{ke}

With use of Equation (2), it can be readily found for the incoming streams of the flamelet that
\begin{eqnarray}
\Big( u_{\xi}- \frac{\omega \chi}{2} \Big)^2 + \Big( u_{\chi}+ \frac{\omega \xi}{2} \Big)^2 +w^2 = u^2 + v^2 +w^2
\end{eqnarray}
Specifically, if account is taken of the moving $\xi, \chi$ coordinates, we see that the same kinetic energy is implied for both frames of reference; although, explicit appearance of the rotational portion of that energy is not given in the rotating frame.

For the constant-density inflow portions of the counterflow structure, we can write for the inflow
\begin{eqnarray}
    u_{\xi} &=& \frac{\partial u_{\xi}}{\partial \xi} \xi   \;\; ; \;\;  u_{\chi} = \frac{\partial u_{\chi}}{\partial \chi} \chi   \;\; ; \;\;  w = \frac{\partial w}{\partial z} z      \nonumber \\
    \frac{u^2 + v^2 +w^2}{2} &=&  \frac{1}{2}\Big( \frac{\partial u_{\xi}}{\partial \xi} \Big)^2 \xi^2  + 
     \frac{1}{2}\Big( \frac{\partial u_{\chi}}{\partial \chi}   \Big)^2  \chi^2
     +   \frac{1}{2}\Big( \frac{\partial w}{\partial z}   \Big)^2  z^2    \nonumber \\
     && +  \frac{1}{8}\omega^2(\xi^2 + \chi^2)  +   \Big(\frac{\partial u_{\chi}}{\partial \chi} -  \frac{\partial u_{\xi}}{\partial \xi}  \Big) \frac{\omega \xi \chi}{2} 
     \label{26_1}
\end{eqnarray}
The right side of Equation (\ref{26_1})
requires knowledge of three quantities for rotational flamelet theory: $S^*, S_1, $ and vorticity $\omega$. The Kolmogorov scale can be used to estimate the lengths.  The kinetic energy per unit mass here is based on Kolmogorov scale velocity fluctuations; so, the value will be less than the statistical value for turbulence kinetic energy $k$. If a relation between the two kinetic energies is provided, knowledge of $k$ provides a relationship between those three quantities.

The turbulence kinetic energy cascade is classically described by assuming the rate of turbulence kinetic energy transfer from any scale to a lower scale is identical.  That means $\epsilon \approx (u')^2/ \tau' \approx (u')^3/ l'$, where $u', \tau'$  and $ l'$  are the associated velocity, time, and length scales for a given portion of the turbulence spectrum. At the Kolmogorov scale (i.e., the smallest scales), those scales are $u_{\kappa}, \tau_{\kappa}$, and $\kappa$, respectively.  At that Kolmogorov scale, the Reynolds number $Re = u_{\kappa} \kappa / \nu \approx 1$ and the strain rate   $u_\kappa/ \kappa \approx \epsilon / u_{\kappa}^2$ is the largest value across all scales.

We will use the relations from Pope, Page 185 \citep{Pope2000}; the Kolmogorov velocity $u_{\kappa} \approx (\epsilon \nu)^{1/4}$ and the Kolmogorov length $\kappa \approx (\nu^3/\epsilon)^{1/4}$. We will average kinetic energy over  a volume equal to $\kappa^3$. The last term in Equation (\ref{26_1}) shows antisymmetry, taking positive values in two quadrants and negative values of identical magnitudes in the other two quadrants of any $\xi,\chi$ plane. Thereby, they have no consequence for our averaging.
Equation (\ref{26_1}) now yields after averaging the right side and relating the left side to $\epsilon$,
\begin{eqnarray}
C_{ke}(\epsilon \nu)^{1/2}   &=& S^{*2}\frac{\nu^{3/2}}{\epsilon^{1/2}}\Big[S_1^2 + 1 + S_2^2  
+ 2 \Big(\frac{\omega}{2 S^*}\Big)^2 \Big]   \nonumber \\
C_{ke}\frac{\epsilon}{\nu} & =& 2S^{*2}\Big[S_1^2 + 1   -  S_1  
+  \Big(\frac{\omega}{2 S^*}\Big)^2   \Big]
\label{26}
\end{eqnarray}
$C_{ke}$ is a nondimensional coefficient of $O(1)$ that appears in the averaging process for the kinetic energy term on the left side of Equations (\ref{26_1}). It accounts for both approximations made in averaging and adjusted decrease in kinetic energy at the Kolmogorov scale because dissipation has occurred over a range of the smallest scales. Realize that because the small-scale time scales are so short compared to filtered scale times, the associated averaging does not reduce final resolution. Thereby, Equations (\ref{23}) and (\ref{26}) have potential to be used with knowledge of  $\epsilon$ from resolved-scale computation   to determine the needed inflow parameters $S^*$ and $\omega$ for the rotational flamelet model.

\section{Scaling of Velocity Gradients}  \label{grad}

Now, we have a relation where, given the value of $\epsilon/\nu$  and a statistical estimate for $S_1$, we can relate $S^*$ to $\omega$. Together with Equation (21), we  have two independent relations for determining $S^*$ and $\omega$ for the rotational flamelet given the resolved scale RANS solutions. Combination of Equations (21) and (24) yields a more direct equation for determining $\omega$. 
\begin{eqnarray}
  \frac{\omega^2}{2 }
  = \Big(C_{ke} - \frac{1}{2}C_{vd}\Big) \frac{\epsilon}{\nu}
  \label{25}
\end{eqnarray}
Interestingly, it is a quadratic equation that  yields two solutions for $\omega$ of equal magnitude and opposite direction; the direction of rotation is irrelevant for major issues.  The magnitudes of $S_1$ and $S_2$ are also irrelevant to the magnitude of $\omega$. The vorticity is of the same order of magnitude as the applied compressive strain rate. The value of the enstrophy (i.e., $\omega^2$) will determine whether the pressure Laplacian from Equation (19) is negative, which is needed to allow the counterflow flamelet to exist. Specifically,
\begin{eqnarray}
    \frac{1}{\rho}\frac{\partial^2 p}{\partial x_i^2} &=& 
    \frac{\omega^2}{2} -\frac{\Phi}{2\mu } =  \Big(C_{ke} - \frac{1}{2}C_{vd}\Big)\frac{\epsilon}{\nu } - C_{vd}\frac{\epsilon}{2\nu}   =  \Big(C_{ke} - C_{vd}\Big)\frac{\epsilon}{\nu }   
\end{eqnarray}
It is required that $C_{ke} < C_{vd}$ in order for the counterflow to exist with pressure peaking at a stagnation point. Furthermore, we must have a positive enstrophy in Equation (\ref{25}). Thus, the complete requirement is $C_{vd}/2 < C_{ke} < C_{vd}$. A given value of $\epsilon$ will produce one acceptable $\omega$ value and one acceptable $S^*$ value. Thereby, we expect one acceptable SDR$_{max}$ value. However, that value cannot be captured by knowing $S^*$ alone.

Thus,  $\epsilon$ is the meaningful tracking variable. From the knowledge of $\epsilon$ and fluid properties, with a choice for the parameter $S_1$, the values of $S^*$, dimensional $ \omega,$ the pressure Laplacian, and the viscous dissipation  follow. Specifically,
\begin{eqnarray}
  S^* = \frac{1}{2}\sqrt{ \frac{C_{vd}\; \epsilon}{ \nu[S_1^2 + 1 - S_1]}  } \;\; &;&  \;\; 
  \omega = \sqrt{ \frac{2[C_{ke} - C_{vd}/2] \;\epsilon}{ \nu}  } \;\; ; \nonumber \\
  \frac{1}{\rho }\frac{\partial^2 p}{\partial x_i^2} = \Big(C_{ke} - C_{vd}\Big)\frac{\epsilon}{\nu } \;\;  &;&  \;\;  \frac{\Phi}{\mu} = C_{vd}\frac{\epsilon}{\nu} 
  \label{29}
\end{eqnarray}

We must be sure that, with any application of the flamelet model for Kolmogorov or near-Kolmogorov scales, $C_{vd}/2 <  C_{ke} < C_{vd}$ is at least supportable in a statistical sense. No  counterflow-based flamelet model can be justified with a positive pressure Laplacian. 

It can be shown that
\begin{eqnarray}
    \frac{\omega^2}{2 S^{*2}}& =&  \frac{[C_{ke} - C_{vd}/2] \;\epsilon}{ \nu S^{*2}}  = 2\frac{[2C_{ke} - C_{vd}][S_1^2 +1 -S_1]}{C_{vd}}
\end{eqnarray}

$\epsilon$ should not be considered a ``progress variable" since its value can increase or decrease in any spacial direction or in time.  We should think in terms of a ``coupling variable" rather than a progress variable.

\section{Flamelet Connection with $\epsilon$} \label{connect}

In this section, the mapping of flamelet solutions to the turbulent kinetic energy dissipation rate $\epsilon$ is presented for two fuel–oxidizer combinations: $\mathrm{H_2/N_2-O_2}$ and JP-5–air. For the $\mathrm{H_2/N_2-O_2}$ configuration, the oxidizer stream consists of pure $\mathrm{O_2}$, while the fuel stream is an equimolar mixture of $\mathrm{H_2}$ and $\mathrm{N_2}$. Both streams enter at 300 K, and the background pressure is set to 10 atm. Gas-phase thermochemistry and transport properties for the $\mathrm{H_2-O_2}$ case are computed using a nine-species skeletal reduction of Version 1.0 of the Foundational Fuel Chemistry Model (FFCM-1) \cite{Smith2016,TAO201818}, where nitrogen is treated as inert. For the JP-5–air configuration at 20 atm, the HyChem A3 skeletal reaction mechanism, which includes 40 chemical species, is employed \cite{wang_physics-based_2018,xu_physics-based_2018}.

Three flamelet models are evaluated: a steady, classical mixture-fraction-based flamelet model, and two variations of the Rotational Flamelet Model (RFM), including models with and without vorticity. All models assume Fickian diffusion at unity Lewis number. Although unsteady flamelet formulations are more suitable for capturing transient phenomena such as ignition and extinction \cite{sirignano_unsteady_2024}, the present analysis focuses on steady-state formulations to the viability of using $\epsilon$ as a coupling variable. The conclusions drawn here remain applicable in the unsteady context.

The classical flamelet solutions are computed in mixture fraction space using the steady laminar flamelet equations \cite{Peters,Peters1984}, with mixture fraction denoted by $Z$. In contrast to the RFM, the momentum equation is not explicitly solved in the classical model. Instead, its effects are introduced by assuming the standard scalar dissipation rate profile
\begin{equation}
\chi(Z) = \frac{2S^*}{\pi}\exp\left(-2\left[\mathrm{erfc}^{-1}(2Z)\right]^2\right)
\end{equation}
This slightly differs from Peters \cite{Peters} by a factor of the square root of the ratio of the two densities for the incoming streams, because, we use the strain rate of the incoming fuel stream rather than the oxidizer stream. As the classical formulation does not include a momentum equation, vorticity effects are inherently excluded. These solutions are generated using the FlameMaster code \cite{flamemaster}.

The remaining two solution sets correspond to Rotational Flamelet Models, generated using an in-house MATLAB solver developed in previous work \cite{Hellwig2025a,Hellwig2025b}. To ensure meaningful comparison with the CFM, the RFM calculations employ equal transverse strain rates ($S_1 = S_2$), with $S_1 = 1/2$ to replicate the two-dimensional axisymmetric counterflow configuration characteristic of the classical flamelet formulation. Naturally, the best comparison with the CFM is for the zero vorticity RFM cases.

The analysis focuses on three key output quantities: the maximum flamelet temperature, the integrated heat release rate (burning rate), and the stoichiometric scalar dissipation rate. These quantities are evaluated as functions of both the inflow strain rate $S^*$ and the turbulent kinetic energy dissipation rate $\epsilon$. The objective here is not to assess whether the rotational flamelet model is superior to the classical flamelet formulation. Results from the classical model are included primarily for reference, as it remains more familiar to many readers. Rather, the focus is on demonstrating how the turbulent kinetic energy dissipation rate $\epsilon$ may serve as a viable coupling variable in both modeling frameworks, and on highlighting how its role differs (particularly with respect to the treatment of vorticity) in the classical versus rotational flamelet approaches.

\begin{figure}[]
  \centering
  \includegraphics[width=3.25in]{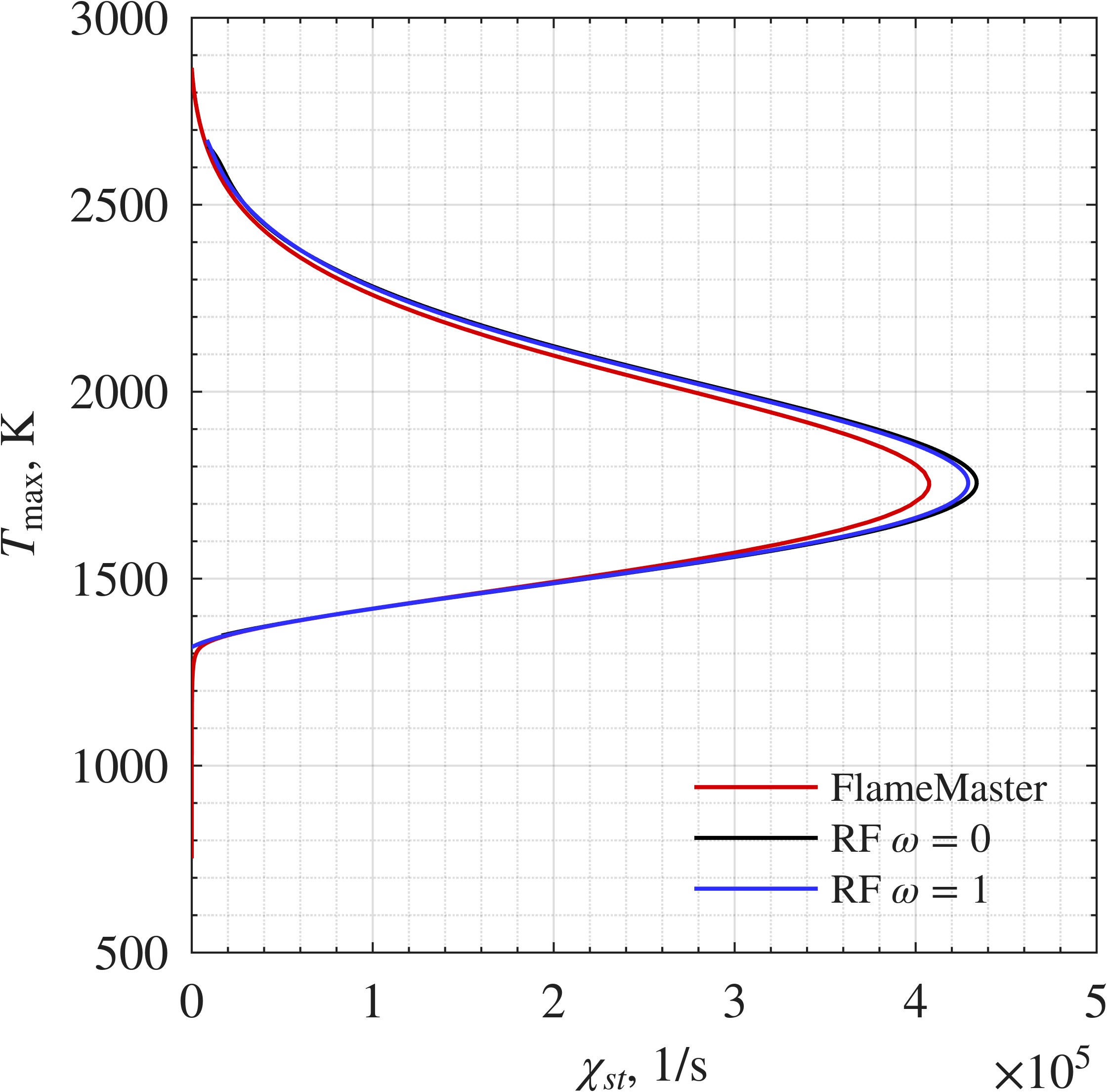}
  \caption{Comparison of three flamelet models for $\mathrm{H_2/N_2-O_2}$ diffusion flames with $T_{max}$ as a function of stoichiometric SDR value ($\chi_{st}$): red, classical result from FlameMaster; black, Rotational Flamelet with zero vorticity; blue, Rotational Flamelet with  vorticity.}
  \label{Tmaxchi}
\end{figure}

Figure \ref{Tmaxchi} presents the maximum flamelet temperature as a function of the stoichiometric scalar dissipation rate, commonly known as S-shaped curves. This figure serves as a common reference for readers familiar with flamelet modeling based on the classical mixture fraction formulation. Differences between the classical and rotational models are evident and anticipated, reflecting the variations in their formulations. The rotational models demonstrate marginally extended flammability limits for both vorticity values. While this presentation, which parametrizes the flame state using scalar dissipation rate, may suggest that vorticity has a negligible impact, it will be shown later that relying on $\chi(Z)$ to describe the flame state is misleading and obscures the significant implications of accounting for vorticity.

\begin{figure}[]
  \centering
  \includegraphics[width=3.25in]{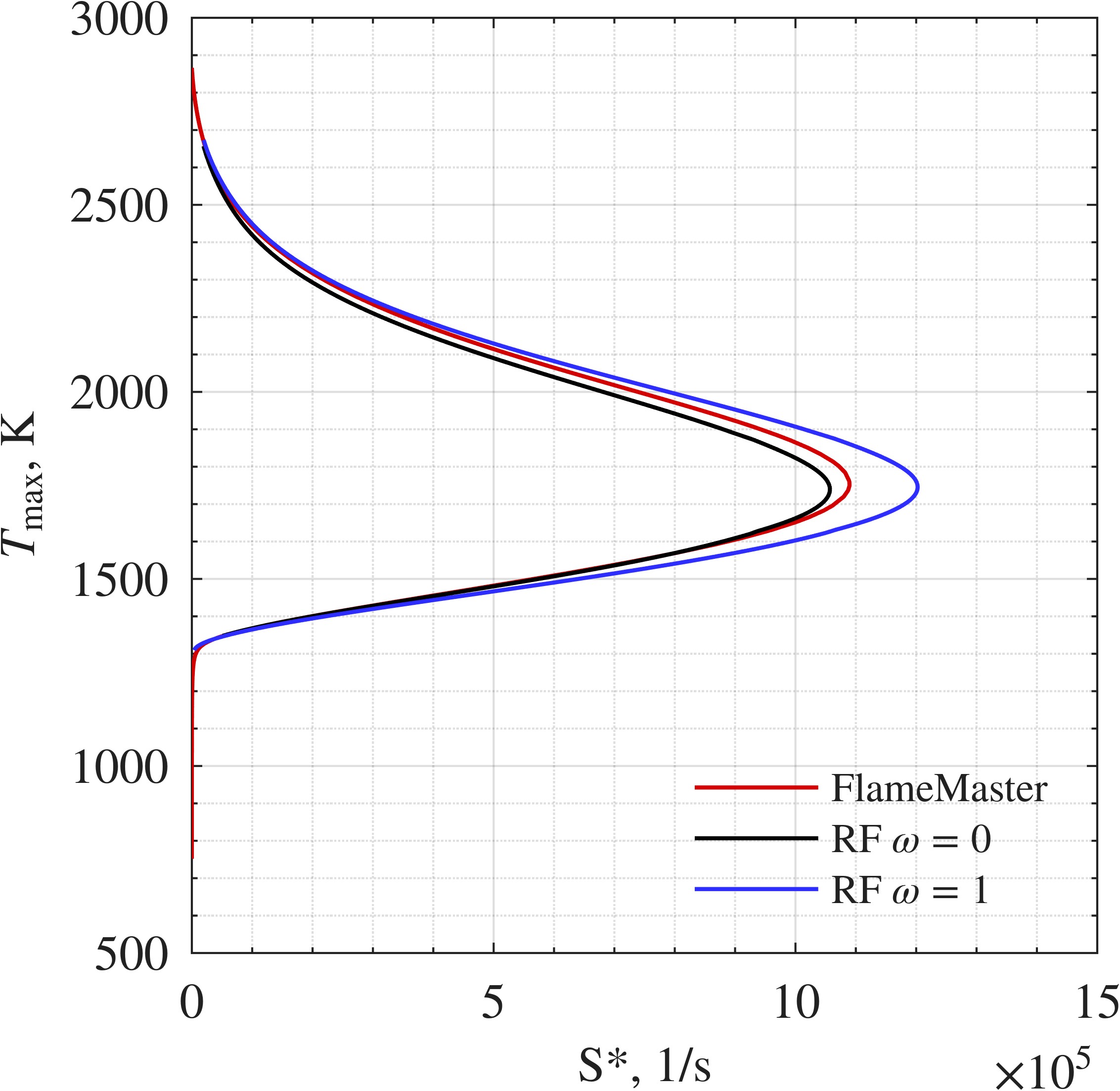}
  \caption{Comparison of three flamelet models for $\mathrm{H_2/N_2-O_2}$ diffusion flames with $T_{max}$ as a function of applied compressive strain rate $S^*$: red, classical result from FlameMaster; black, Rotational Flamelet with zero vorticity; blue, Rotational Flamelet with  vorticity.  }
  \label{TmaxS}
\end{figure}

Figure \ref{TmaxS} displays the same S-shaped curves, but now parameterized by the inflow strain rate $S^*$. For the classical model, the S-shaped curve plotted against strain rate remains qualitatively similar to that in Figure \ref{Tmaxchi}. This occurs because $S^*$ is related to $\chi_{st}$ by a constant through assumed functional form of $\chi(Z)$. However, in the case of the rotational models (where $\chi(Z)$ is not assumed but instead determined by solving the momentum equation) the effect of vorticity on the flamelet becomes more pronounced. The zero-vorticity case closely resembles the classical flamelet model, which is expected, as the absence of vorticity aligns with the classical formulation where vorticity is not considered. However, when vorticity is included, a significant increase in the extinction inflow strain rate is observed.

\begin{figure}[]
  \centering
  \includegraphics[width=3.25in]{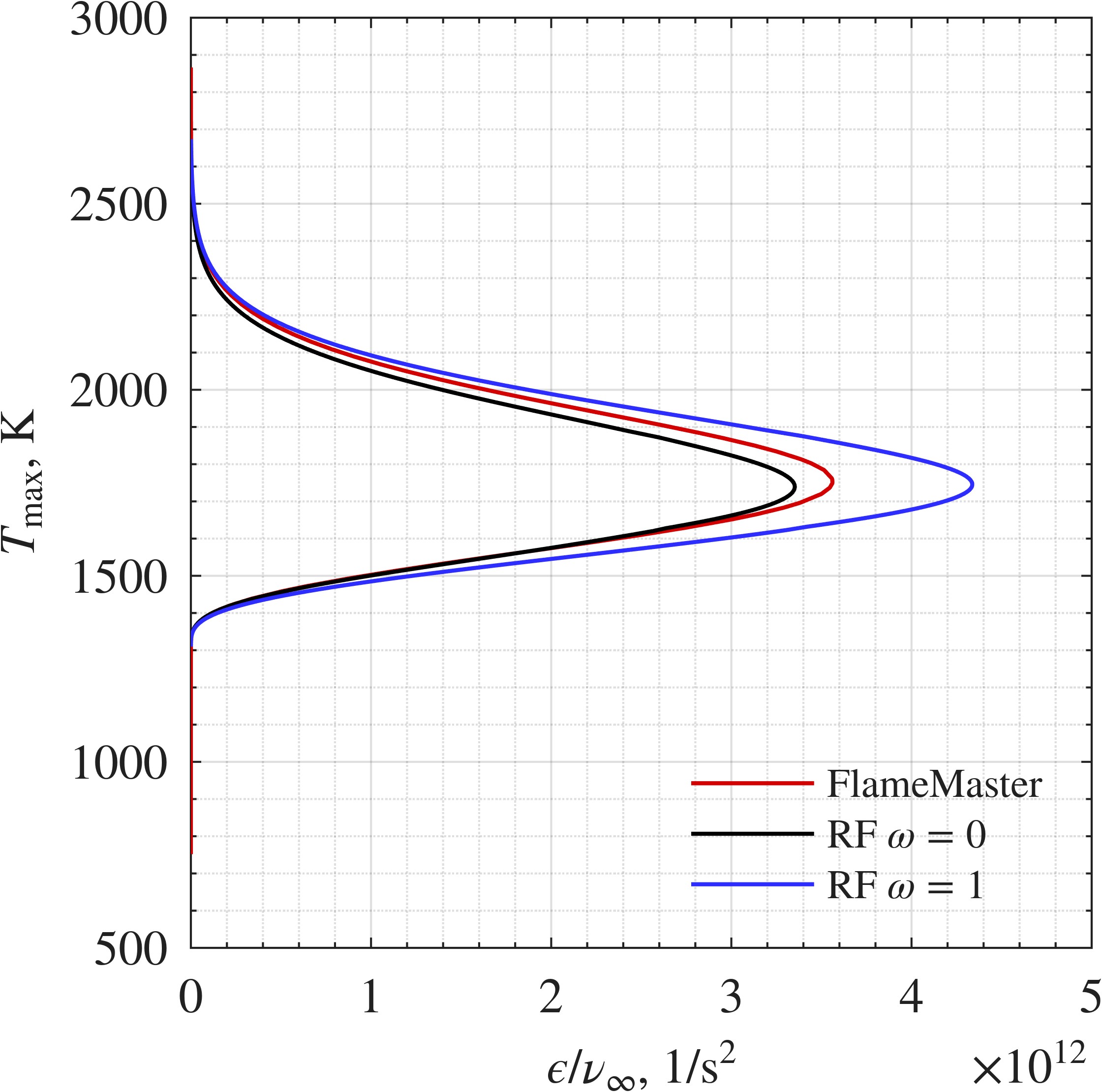}
  \caption{Comparison of three flamelet models for $\mathrm{H_2/N_2-O_2}$ diffusion flames with $T_{max}$ as a function of $\epsilon / \nu_{\infty}$: red, classical result from FlameMaster; black, Rotational Flamelet with zero vorticity; blue, Rotational Flamelet with vorticity. }
  \label{Tmaxeps}
\end{figure}

In Figure \ref{Tmaxeps}, the maximum temperature is parameterized by $\epsilon/\nu_{\infty}$. To achieve this, $S^*$ is related to $\epsilon/\nu$ using the first expression from Equation \ref{29}. Although the value of $C_{vd}$ is not known at this time, a value of 1 has been selected for demonstration purposes. With $S_1=1/2$, the inflow strain rate becomes $S^*=\sqrt{\epsilon/(3\nu)}$. This illustrates how $\epsilon$, which is determined from the resolved scale, can serve as a tracking  effectively connecting the flamelet to resolved scale computations.

\begin{figure}[]
  \centering
  \includegraphics[width=3.25in]{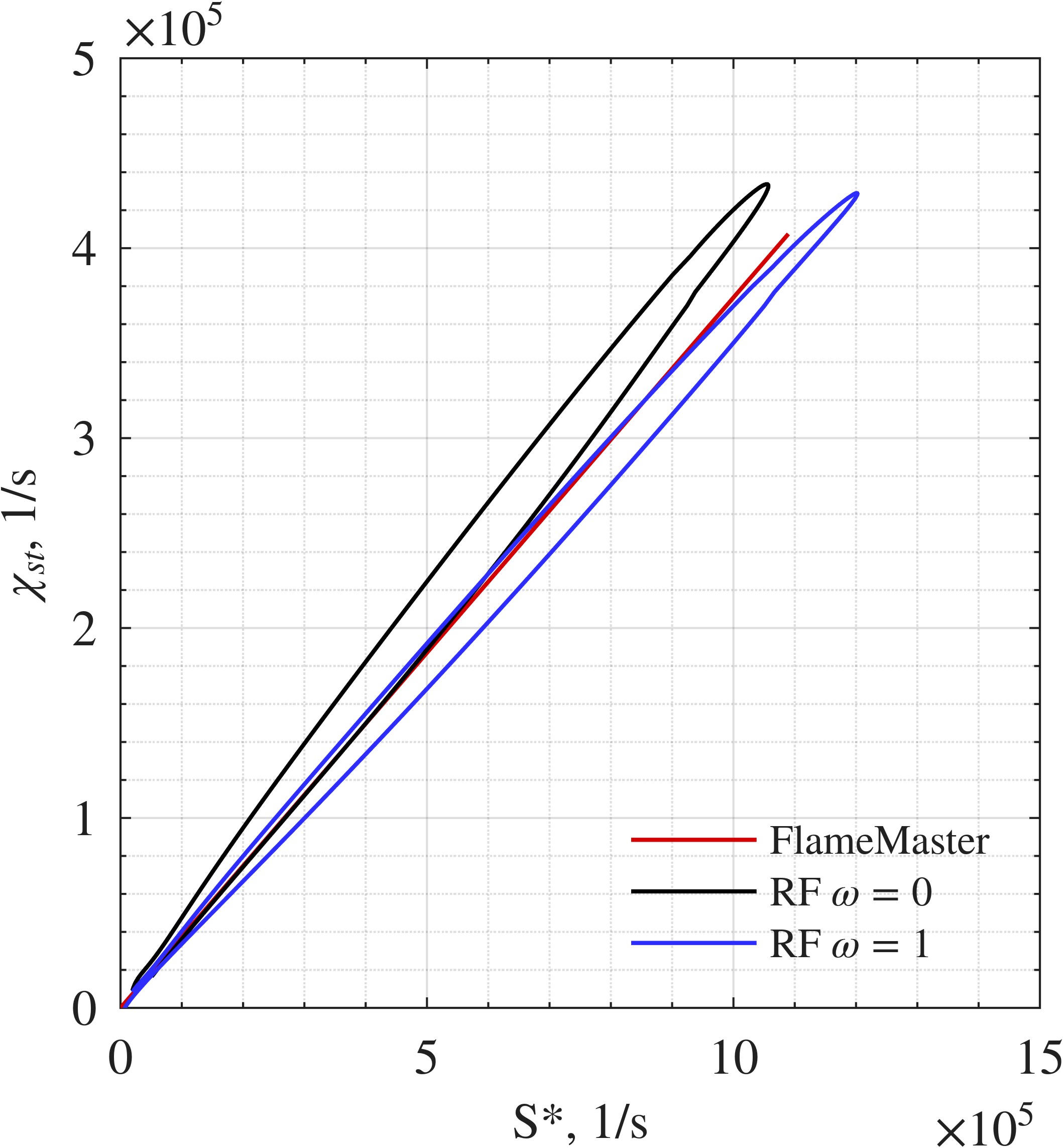}
  \caption{Comparison of three flamelet models for $\mathrm{H_2/N_2-O_2}$ diffusion flames with $\chi_{st}$ as a function of $S^*$: red, classical result from FlameMaster; blue, Rotational Flamelet with zero vorticity; black, Rotational Flamelet with vorticity.}
  \label{chistS}
\end{figure}

\begin{figure}[]
  \centering
  \includegraphics[width=3.25in]{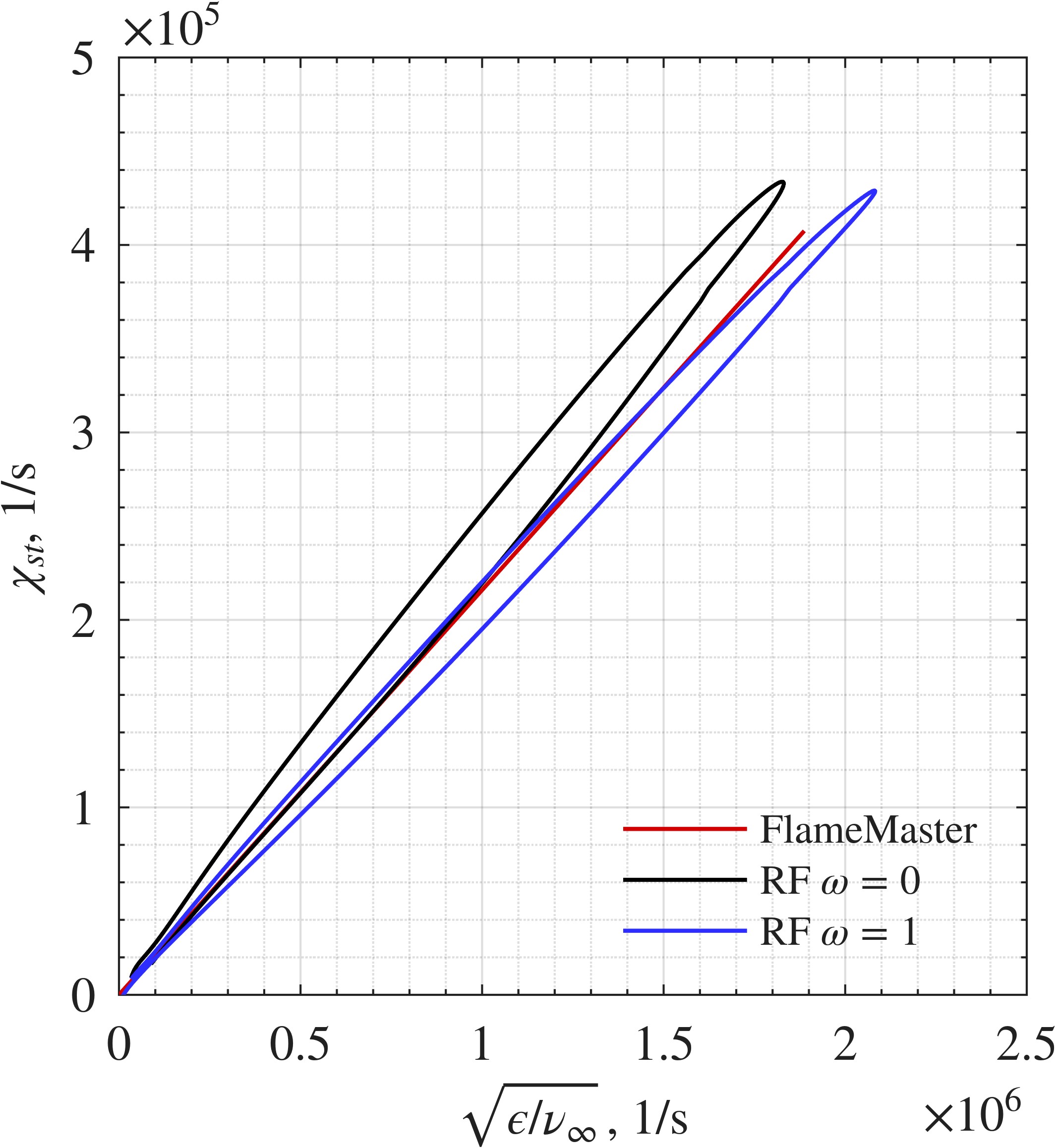}
  \caption{Comparison of three flamelet models for $\mathrm{H_2/N_2-O_2}$ diffusion flames with $\chi_{st}$ as a function of $\sqrt{\epsilon / \nu_{\infty}}$: red, classical result from FlameMaster; black, Rotational Flamelet with zero vorticity; blue, Rotational Flamelet with vorticity.}
  \label{chistSqrteps}
\end{figure}

Figures \ref{chistS} and \ref{chistSqrteps} plot stoichiometric scalar dissipation rates $\chi_{st}$ for the three models against $S^*$ and $\sqrt{\epsilon/\nu_{\infty}}$, respectively. The $\chi_{st}$ used in the classical model, represented by the red line, is bijective and grows linearly with $S^*$, as a result of the assumed functional form of $\chi(Z,\omega)$. As previously discussed, in the Rotational Flamelet model, scalar dissipation is not assumed but determined by solving the momentum equation. The differences between the blue and black curves clearly demonstrate how scalar dissipation rate becomes a function of vorticity, i.e., $\chi(Z,\omega)$. Furthermore, contrary to the classical model, stable and unstable solutions have distinct $\chi_{st}$ values. 

This dependence of $\chi(Z)$ on $\omega$ has significant implications. When viewing these figures alongside Figure \ref{Tmaxchi}, it becomes evident that, for a given value of $\epsilon$ determined from the resolved scale, which imposes a corresponding $S^*$ value, different vorticity values result in two distinct $\chi_{st}$, values, leading to two different flamelet solutions along the S-shaped curve. These solutions exhibit different maximum temperatures and integrated burning rates, regardless of the fact that the S-shaped curves for different vorticity values collapse onto each other when parameterized by $\chi_{st}$. This clearly shows that a flamelet solution cannot be determined based on $S^*$ alone.

Figure \ref{chistSqrteps} shows the same $\chi_{st}$ curves plotted against the square root of $\epsilon/\nu_{\infty}$. $\epsilon$ is the rate of dissipation of k, the turbulence kinetic energy per unit mass. It follows that the magnitude of velocity fluctuation decays at a rate proportional to $\sqrt{\epsilon}$. If we assume that a scalar fluctuation decreases proportionately to velocity and $\chi(Z)$ represents a rate of dissipation for a scalar fluctuation, the proportionality of $\chi(Z)$ and $\sqrt{\epsilon}$ are observed and expected.

\begin{figure}[]
  \centering
  \includegraphics[width=3.25in]{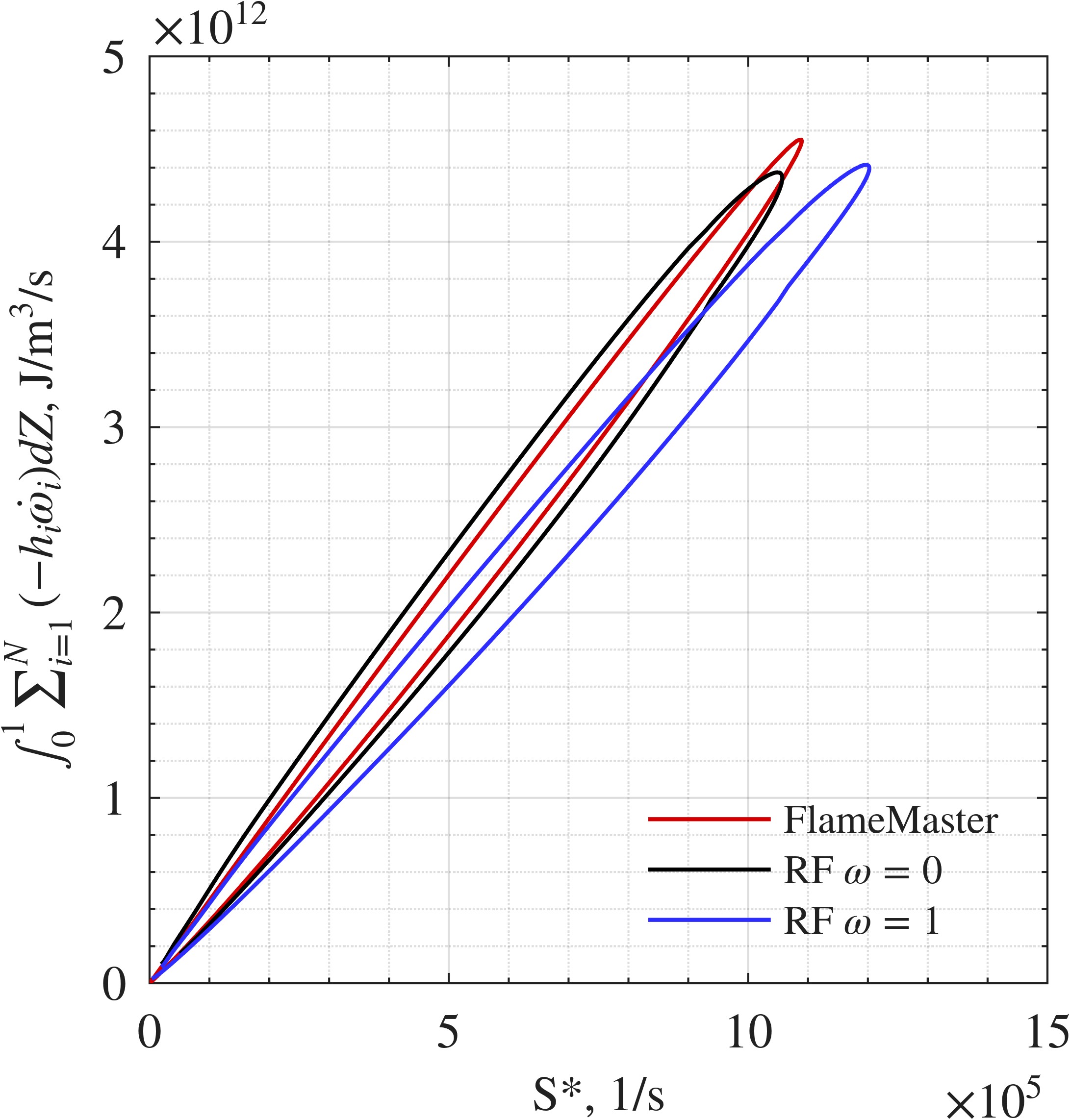}
  \caption{Comparison of three flamelet models for $\mathrm{H_2/N_2-O_2}$ diffusion flames with integrated burning rate as a function of $S^*$: red, classical result from FlameMaster; black, Rotational Flamelet with zero vorticity; blue, Rotational Flamelet with vorticity.}
  \label{IBRsstar}
\end{figure}

\begin{figure}[]
  \centering
  \includegraphics[width=3.25in]{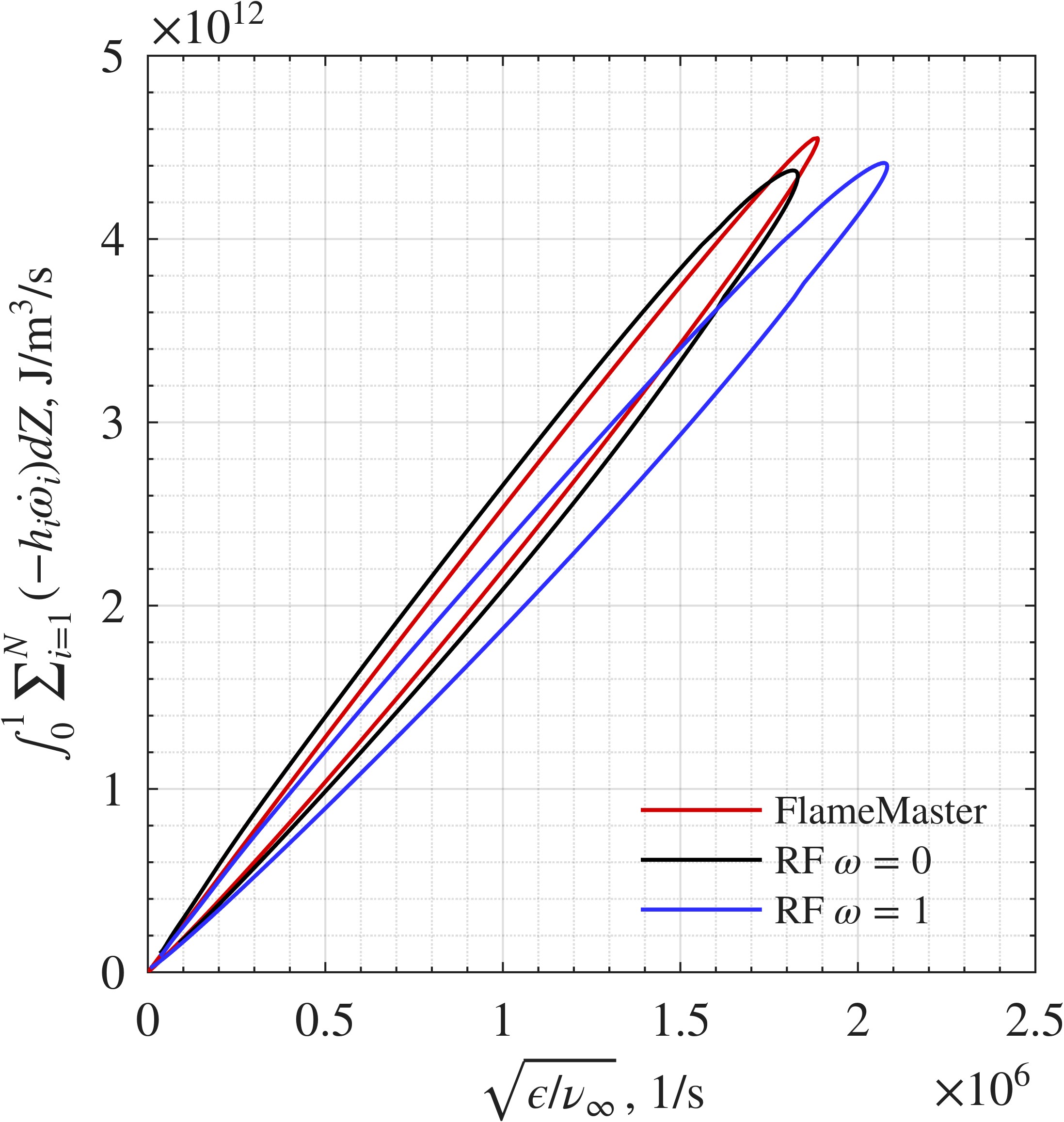}
  \caption{Comparison of three flamelet models for $\mathrm{H_2/N_2-O_2}$ diffusion flames with integrated burning rate as a function of $\sqrt{\epsilon / \nu_{\infty}}$: red, classical result from FlameMaster; black, Rotational Flamelet with zero vorticity; blue, Rotational Flamelet with vorticity.}
  \label{IBReps}
\end{figure}

More relevant than flame temperature is the heat release, which will in practice be the parameter extracted from the flamelet model and introduced into the resolved-scale computations. Figures \ref{IBRsstar} and \ref{IBReps} depict the integrated burning rate for the three models as functions of $S^*$ and $\sqrt{\epsilon/\nu_{\infty}}$, respectively. Here, the integrated burning rate is calculated as the heat release rate integrated over $Z$. All three models show that the integrated burning rate increases in a nearly linear manner with $S^*$, indicating that it grows as the scale decreases. As expected, the Rotational Flamelet model with zero vorticity most closely resembles the classical model, with differences in integrated burning rates due to distinctions in the underlying formulations of the two models. In the presence of vorticity (blue curve), the integrated burning rate shows a significant decrease compared to the zero-vorticity case.

\begin{figure}[]
  \centering
  \includegraphics[width=3.25in]{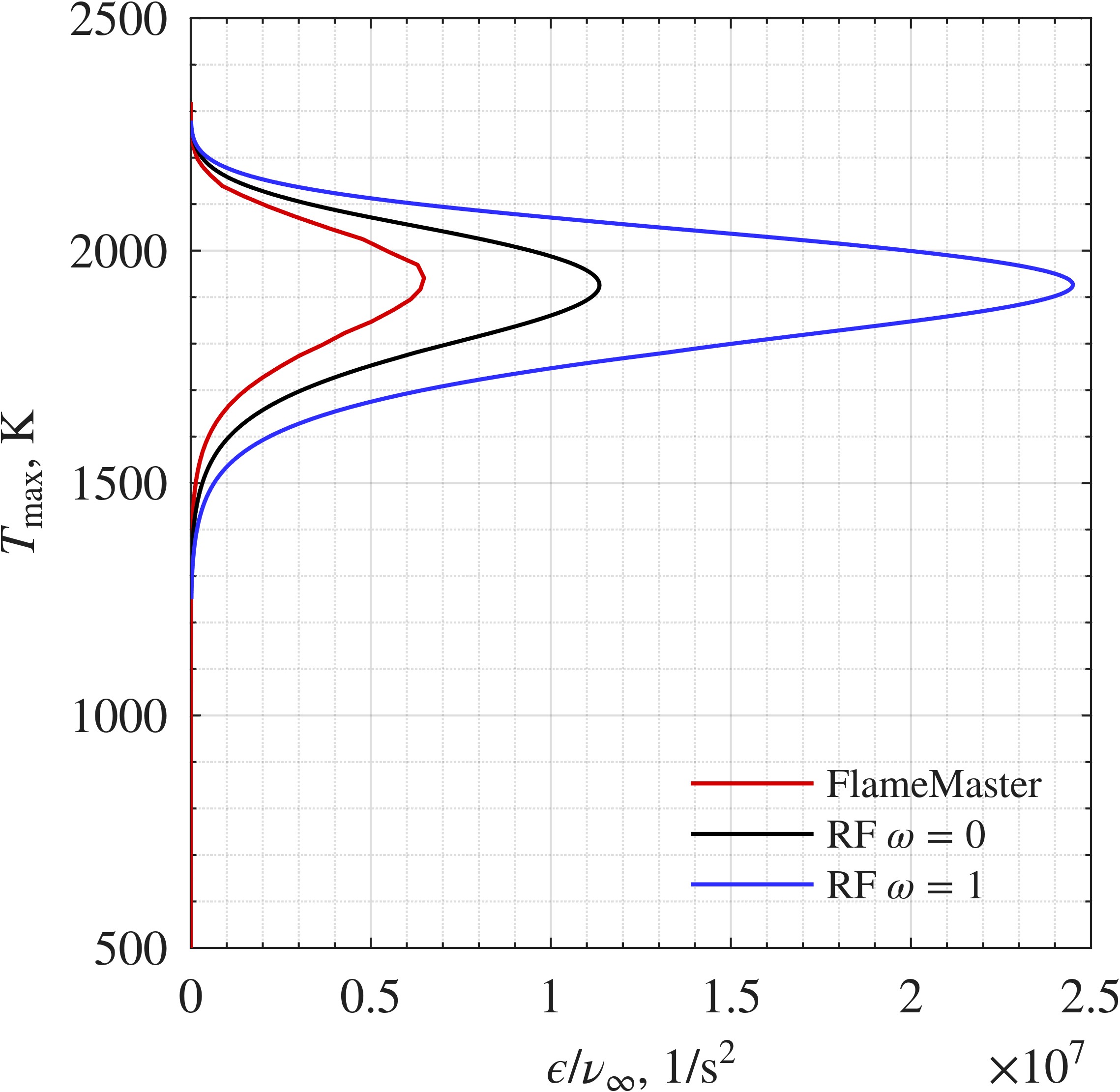}
  \caption{Comparison of three flamelet models for JP-5 diffusion flames with $T_{max}$ as a function of $\epsilon / \nu_{\infty}$: red, classical result from FlameMaster; black, Rotational Flamelet with zero vorticity; blue, Rotational Flamelet with vorticity.   }
  \label{jp5Tmaxeps}
\end{figure}

\begin{figure}[]
  \centering
  \includegraphics[width=3.25in]{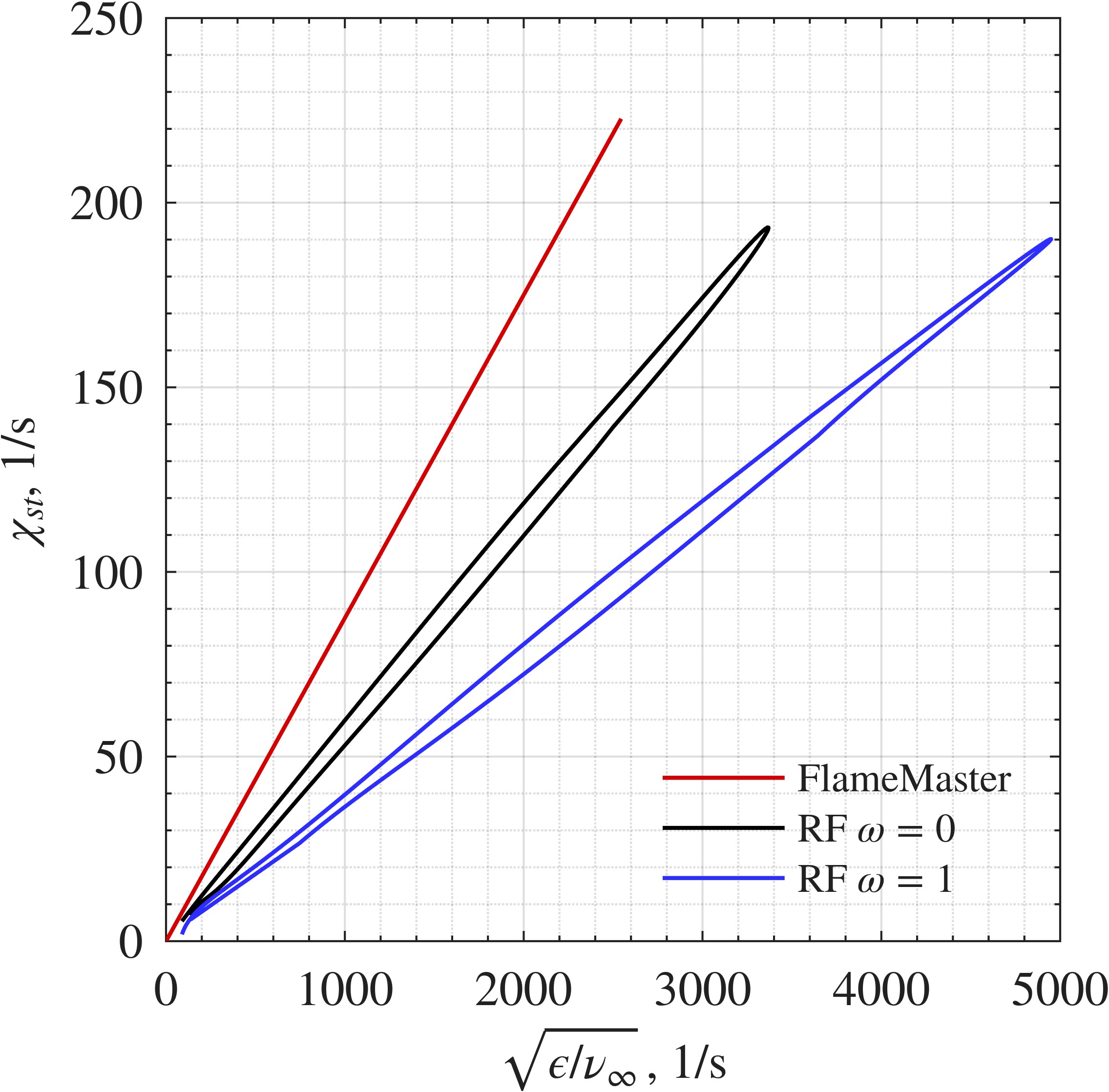}
  \caption{Comparison of three flamelet models for JP-5 diffusion flames with $\chi_{st}$ as a function of $\sqrt{\epsilon / \nu_{\infty}}$: red, classical result from FlameMaster; black, Rotational Flamelet with zero vorticity; blue, Rotational Flamelet with vorticity.}
  \label{jp5chistSqrteps}
\end{figure}

\begin{figure}[]
  \centering
  \includegraphics[width=3.25in]{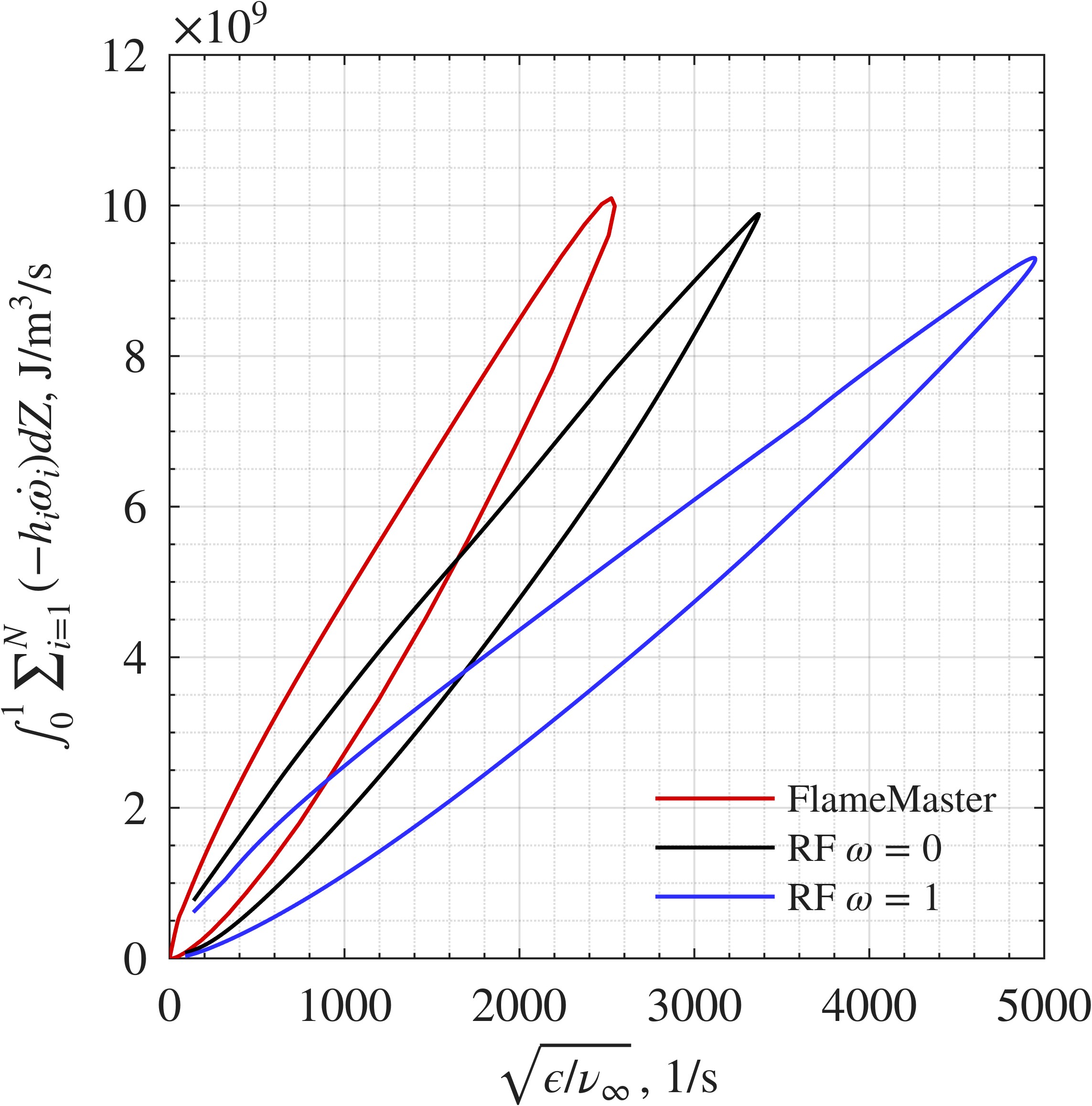}
  \caption{Comparison of three flamelet models for JP-5 diffusion flames with integrated burning rate as a function of $\sqrt{\epsilon / \nu_{\infty}}$: red, classical result from FlameMaster; black, Rotational Flamelet with zero vorticity; blue, Rotational Flamelet vorticity.}
  \label{jp5IBReps}
\end{figure}

Figures \ref{jp5Tmaxeps}, \ref{jp5chistSqrteps}, and \ref{jp5IBReps} show key results for the jet fuel JP-5 burning with air. The JP-5 air models are generated for an oxidizer stream consisting of pure air and a fuel stream consisting of pure JP-5. Both streams enter at 300 K, with a background pressure of 20 atm. Gas-phase kinetics, thermal properties, and transport properties are calculated using the A3 skeletal version of the HyChem Combustion reaction models \cite{wang_physics-based_2018,xu_physics-based_2018}. Nitrogen is treated as an inert species. We see that JP-5 with its much slower chemical reaction rate than hydrogen will not sustain as high  a strain rate; the flammability domain is lower by several orders of magnitude for any variable of interest $S^*$, $\epsilon$, or $\sqrt{\epsilon}$.  On the other hand, the percentage increase of flammability limit due to vorticity is substantially greater. 

\begin{figure}[]
  \centering
  \includegraphics[width=3.25in]{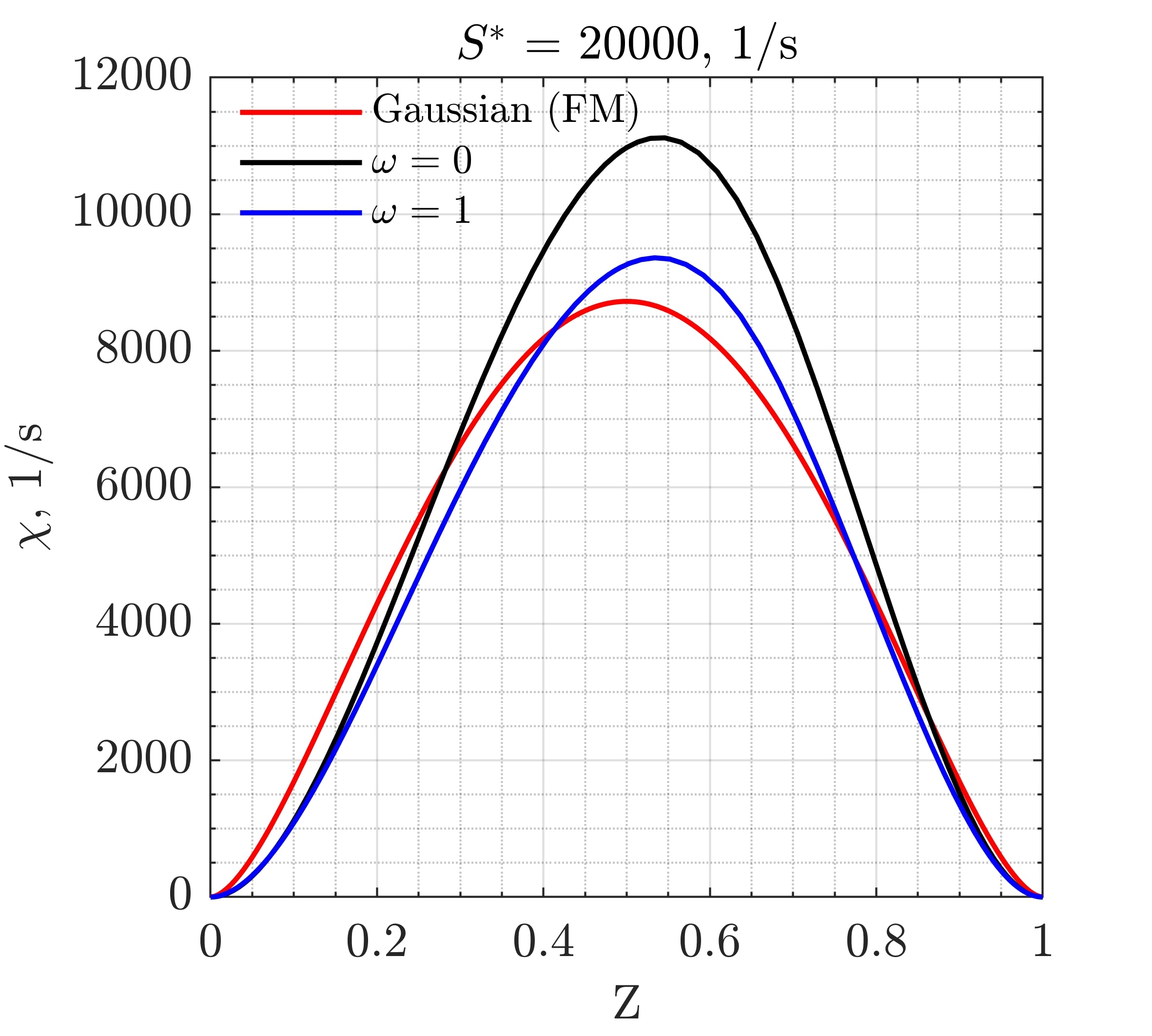}
  \caption{Comparison of three flamelet models for $\mathrm{H_2/N_2-O_2}$ diffusion flames with SDR as a function of mixture fraction $Z$: red, classical from FlameMaster; black, Rotational Flamelet with zero vorticity; blue, Rotational Flamelet with vorticity.}
  \label{Chi}
\end{figure}

\begin{figure}[]
  \centering
  \includegraphics[width=3.25in]{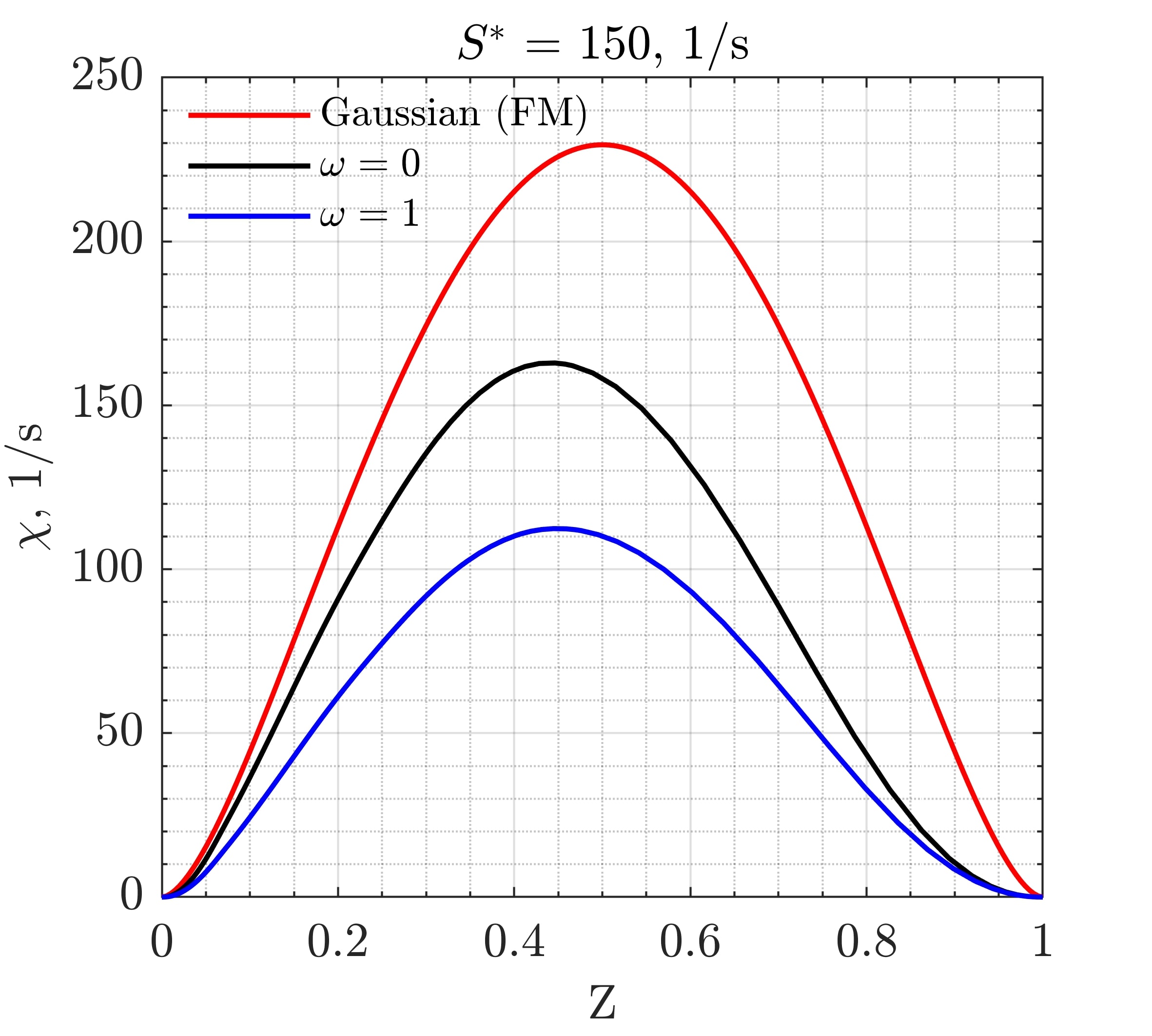}
  \caption{Comparison of three flamelet models for JP-5 $-$ Air diffusion flames with SDR as a function of mixture fraction $Z$: red, classical result from FlameMaster; black, Rotational Flamelet with zero vorticity; blue, Rotational Flamelet with vorticity.   }
  \label{ChiJP5}
\end{figure}

Figures \ref{Chi} and \ref{ChiJP5} are examples of scalar dissipation rate $\chi(Z)$, versus mixture fraction $Z$, for $\mathrm{H_2}, \mathrm{O_2}, \mathrm{N_2}$ combustion and JP-5, air combustion, respectively. The blue and black curves illustrate the difference in scalar dissipation rate created by solving the momentum equations with vorticity in addition to the energy and species continuity equations. Clearly, vorticity has an effect on scalar dissipation rate and will thus alter the solution of the flamelet equations. 

Furthermore, the peak of the scalar dissipation rate curve shows bias towards the stream with higher thermal diffusivity. This is seen from the $\mathrm{H_2}, \mathrm{O_2}, \mathrm{N_2}$ peak sloping toward the fuel stream while the JP-5 peaks slopes away from the fuel stream. For $Le=1$, the mass diffusivity value is set to the thermal diffusivity value which has the product of density and specific heat in the denominator. Both terms in that denominator product minimize on the fuel-rich (fuel-lean) side for $\mathrm{H_2}/\mathrm{N_2-O_2}$ (JP-5 $-$ Air), causing diffusivity to maximize on that side. This results in $dZ/dy$, its square, and the coefficient $D$ maximizing and causing $\chi(Z)$ to maximize on that preferred side. Let us, for the sake only of argument and display here, consider that, in the vicinity of maximum SDR, the velocity $v = -S^*y$ and $D$ is approximately constant over a neighborhood. Treatment of the second-order ODE for $Z(y)$ as a first-order ODE for $dZ/dy$ yields a local behavior where $dZ/dy  \propto \exp{[-S^*y^2/(2D)]}$ and $\chi \propto D \exp{[-S^*y^2/D]}$. The larger $D$ in the denominator of the exponent causes a slower decline from the maximum value and thereby a larger value of the exponential term as well as its coefficient. This difference in thermal diffusivity for unitary Lewis number is due to the relatively high specific heat for JP-5 and $\mathrm{O_2}$ relative to their respective opposing streams.

Since the scalar dissipation rate is proportional to strain rate, we should expect SDR$_{max}/ S^* \propto \epsilon^{1/2}$ and scalar gradients should behave as $\epsilon^{1/4}$.


\section{Conclusions}\label{conclusions}

 The turbulence kinetic energy dissipation rate $\epsilon$ have been defined at the resolved scale of turbulent flow fields, including flows with turbulent combustion. $\epsilon$ is readily defined in both RANS and detached-eddy simulations. It is an excellent parameter for closure with sub-grid non-premixed flamelet models to determine the vorticity $\omega$ and the compressive normal strain rate $S^*$ at the smallest length scales, given the turbulence description at the resolved scale.  Thereby, there is no need for creation of a progress variable or new variables in the resolved scale. 
 The analysis relates $\epsilon$ with the strain rate, vorticity, viscous dissipation rate, scalar gradients, scalar dissipation rate, and burning rate at the smallest turbulence length scales where diffusion-controlled burning is expected to be faster than at larger length scales and thereby dominant.

 Two flamelet inflows with the same $\epsilon$ value at the resolved scale can still have statistical variations with a different pair of $S^*$ and $\omega$ values. So, the coefficients $C_{vd}$ and $C_{ke}$ have been used in the respective relations for $\epsilon$ with Kolmogorov-scale viscous dissipation rate and Kolmogorov-scale kinetic energy. Without further statistical refinement, the theory has the assumption of uniform coefficients over the resolved flow field. Conditions in the center of the flamelet can be identical, even for flamelets with different  values for $\epsilon, S^*,$ and $\omega$. For example, flamelets with the same values for maximum temperature $T_{max}$, scalar dissipation rate at the stoichiometric point in the flow $\chi_{st}$, and / or integrated burning rate can experience different resolved scale values such as $\epsilon$ or inflow values such as $S^*$ or $\omega$.

 Direct numerical simulations in the literature \cite{Nomura1992,Nomura1993,Boratav1996,Boratav1998,Betchov1956} indicate that values for Kolmogorov-scale vorticity and normal strain rate can be expected to have the same order of magnitude. With this consideration, we find that scalar gradients in the flamelet tend to scale with $\epsilon^{1/4}$. This means that scalar dissipation rate scales as $\epsilon^{1/2}$.

A key conclusion here is that a flamelet model that did not consider vorticity would be unable to match key physical parameters determined on the resolved scale.

There are remaining questions. Each pair of values for vorticity and strain rate (at the same value for $\epsilon/ \nu$) produce both a stable and an unstable branch. The appearance of the stability issue is not a mechanical impact but rather a thermal impact that relates to balance between chemical energy conversion rate on one hand and heat transfer rate on the other hand.  

Statistical data on the values of $S_1$ and $S_2$ for reacting flows would be very useful. It likely would come from DNS, but experiments might be useful.

\section*{Acknowledgement(s)}

We appreciate the advice on RANS theory provided by Professor Feng Liu, on combustion theory by Professor Xian Shi, and on turbulence theory by Professors Perry Johnson and Said Elghobashi. Professor Heinz Pitsch is thanked for sharing the Flamemaster code.

\section*{Disclosure statement}

The authors report there are no competing interests to declare.

\section*{Funding}

The effort was supported by AFOSR through Award FA9550--22-1-0191 managed by Dr. Mitat Birkan and now by Dr. Justin Koo and by ONR through Award N00014-22-1-2467 managed by Dr. Steven Martens.

\section*{Notes on contributor(s)}

William Sirignano is a Distinguished Professor at the Samueli School of Engineering
at the University of California, Irvine. His vast research portfolio over the past 60
years includes work on combustion theory, computational methods, fluid dynamics,
multiphase flows, combustion instability, and propulsion systems. His current research focuses on numerical simulation of practical turbulent reacting flows through Reynolds-Averaged Navier Stokes and Large-Eddy Simulations and high-performance upgrades to gas-turbine engines. He received his M.A. degree in 1962 and Ph.D. degree in 1964, both from Princeton University.

Wes Hellwig is a Ph.D. student in Mechanical and Aerospace Engineering at the
Samueli School of Engineering at the University of California, Irvine. His emphasis is in fluid dynamics and propulsion and his research focuses on numerical simulation
of turbulent reacting flows. He received his M.S. degree in 2023 from the University of
California, Irvine.

Sylvain L. Walsh is a Ph.D. student in Mechanical and Aerospace Engineering at the Samueli School of Engineering, University of California, Irvine. His research centers on combustion closure models for RANS and LES simulations. He received his Bachelor’s degree from the Universitat Politècnica de Catalunya in 2021 and his M.S. degree from the University of California, Irvine in 2023.

\bibliographystyle{tfq}
\bibliography{rev2-epsilon_flamelet}

\newpage
\section{Tables}
\begin{table}[h]
\caption{Comparison of flamelet models and coupling procedures }
\label{tab}
\begin{tabular}{ccc}
\hline
\multicolumn{1}{|c|}{\textbf{Model}}                & \multicolumn{1}{c|}{\textbf{Constraint to the flamelet}}                                                                      & \multicolumn{1}{c|}{\textbf{Resolved-scale coupling procedure}} \\ \hline
\multicolumn{1}{|c|}{\multirow{2}{*}{\textbf{CFM}}} & \multicolumn{1}{c|}{Scalar dissipation rate, $\chi$\textsuperscript{\textdagger}}                                                                          & \multicolumn{1}{c|}{PDFs for $Z$ and $\chi$\textsuperscript{\textdagger}}                    \\ \cline{2-3} 
\multicolumn{1}{|c|}{}                              & \multicolumn{1}{c|}{\begin{tabular}[c]{@{}c@{}}$\chi$ mapped to the \\ flamelet inflow strain rate, $S^*$\end{tabular}}       & \multicolumn{1}{c|}{$\epsilon$ for $S^*$ and a PDF for $Z$}     \\ \hline
\multicolumn{1}{|c|}{\textbf{FPV}}                  & \multicolumn{1}{c|}{\begin{tabular}[c]{@{}c@{}}Progress variable $C$\\ (weighted sum of product mass fractions)\textsuperscript{\textdagger}\end{tabular}} & \multicolumn{1}{c|}{PDFs for $Z$ and $C$\textsuperscript{\textdagger}}     \\ \hline
\multicolumn{1}{|c|}{\textbf{RFM}}                  & \multicolumn{1}{c|}{$S^*$ and vorticity $\omega$}                                                                             & \multicolumn{1}{c|}{$\epsilon$ for $S^*$ and $\omega$}          \\ \hline
\multicolumn{3}{l}{\textsuperscript{\textdagger}Conventional approach}                                                                                                                                                                                                            
\end{tabular}
\end{table}

\newpage

\section{Figures}




\newpage
\section{Figure Captions (as list)}
\begin{enumerate}
     \item Transformation to $\xi, \chi, z'$ rotating coordinate system from $x', y', z'$ Newtonian system. $\theta$ increases in the counterclockwise direction.   
     \item Comparison of three flamelet models for $\mathrm{H_2/N_2-O_2}$ diffusion flames with $T_{max}$ as a function of stoichiometric SDR value ($\chi_{st}$): red, classical result from FlameMaster; black, Rotational Flamelet with zero vorticity; blue, Rotational Flamelet with  vorticity.
     \item Comparison of three flamelet models for $\mathrm{H_2/N_2-O_2}$ diffusion flames with $T_{max}$ as a function of applied compressive strain rate $S^*$: red, classical result from FlameMaster; black, Rotational Flamelet with zero vorticity; blue, Rotational Flamelet with  vorticity.  
     \item Comparison of three flamelet models for $\mathrm{H_2/N_2-O_2}$ diffusion flames with $T_{max}$ as a function of $\epsilon / \nu_{\infty}$: red, classical result from FlameMaster; black, Rotational Flamelet with zero vorticity; blue, Rotational Flamelet with vorticity. 
     \item Comparison of three flamelet models for $\mathrm{H_2/N_2-O_2}$ diffusion flames with $\chi_{st}$ as a function of $S^*$: red, classical result from FlameMaster; blue, Rotational Flamelet with zero vorticity; black, Rotational Flamelet with vorticity.
     \item Comparison of three flamelet models for $\mathrm{H_2/N_2-O_2}$ diffusion flames with $\chi_{st}$ as a function of $\sqrt{\epsilon / \nu_{\infty}}$: red, classical result from FlameMaster; black, Rotational Flamelet with zero vorticity; blue, Rotational Flamelet with vorticity.
     \item Comparison of three flamelet models for $\mathrm{H_2/N_2-O_2}$ diffusion flames with integrated burning rate as a function of $S^*$: red, classical result from FlameMaster; black, Rotational Flamelet with zero vorticity; blue, Rotational Flamelet with vorticity.
     \item Comparison of three flamelet models for $\mathrm{H_2/N_2-O_2}$ diffusion flames with integrated burning rate as a function of $\sqrt{\epsilon / \nu_{\infty}}$: red, classical result from FlameMaster; black, Rotational Flamelet with zero vorticity; blue, Rotational Flamelet with vorticity.
     \item Comparison of three flamelet models for JP-5 diffusion flames with $T_{max}$ as a function of $\epsilon / \nu_{\infty}$: red, classical result from FlameMaster; black, Rotational Flamelet with zero vorticity; blue, Rotational Flamelet with vorticity.
     \item Comparison of three flamelet models for JP-5 diffusion flames with $\chi_{st}$ as a function of $\sqrt{\epsilon / \nu_{\infty}}$: red, classical result from FlameMaster; black, Rotational Flamelet with zero vorticity; blue, Rotational Flamelet with vorticity.
     \item Comparison of three flamelet models for JP-5 diffusion flames with integrated burning rate as a function of $\sqrt{\epsilon / \nu_{\infty}}$: red, classical result from FlameMaster; black, Rotational Flamelet with zero vorticity; blue, Rotational Flamelet vorticity.
     \item Comparison of three flamelet models for $\mathrm{H_2/N_2-O_2}$ diffusion flames with SDR as a function of mixture fraction $Z$: red, classical from FlameMaster; black, Rotational Flamelet with zero vorticity; blue, Rotational Flamelet with vorticity.
     \item Comparison of three flamelet models for JP-5 $-$ Air diffusion flames with SDR as a function of mixture fraction $Z$: red, classical result from FlameMaster; black, Rotational Flamelet with zero vorticity; blue, Rotational Flamelet with vorticity.

\end{enumerate}

\end{document}